\documentclass[useAMS,usenatbib,fleqn]{mnras}
\usepackage{graphicx,subfig,lscape}
\usepackage[usenames,dvipsnames]{xcolor}
\usepackage{newtxtext,newtxmath}
\usepackage[T1]{fontenc}

\newcommand{\kms}{km\,s$^{-1}$}
\newcommand{\vs}{$v_{\rm e} \sin i$}
\newcommand{\hd}{HD\,66051}
\newcommand{\bz}{$\langle B_{\rm z} \rangle$}

\defcitealias{niemczura:2017}{N17}

\newcommand{\figps}[1]{\resizebox{\hsize}{!}{\rotatebox{0}{\includegraphics{#1}}}}

\newcommand{\firps}[1]{\resizebox{\hsize}{!}{\rotatebox{-90}{\includegraphics{#1}}}}

\newcommand{\fifps}[2]{\centering\resizebox{#1}{!}{\includegraphics{#2}}}

\newcommand{\fps}[3]{\resizebox{#1}{!}{\rotatebox{#2}{\includegraphics{#3}}}}

\title[Magnetic eclipsing binary HD\,66051]{HD\,66051: the first eclipsing binary hosting an early-type magnetic star%
\thanks{Based on observations obtained at the Canada-France-Hawaii Telescope (CFHT), which is operated by the National Research Council of Canada, the Institut National des Sciences de l'Univers of the Centre National de la Recherche Scientifique of France, and the University of Hawaii.}}

\author[O. Kochukhov et al.]{
O. Kochukhov$^{1}$\thanks{E-mail: oleg.kochukhov@physics.uu.se},
C. Johnston$^{2}$,
E. Alecian$^{3}$, 
G.~A.~Wade$^{4}$,
and the BinaMIcS collaboration\\
$^{1}$Department of Physics and Astronomy, Uppsala University, 751 20 Uppsala, Sweden\\
$^{2}$Instituut voor Sterrenkunde, KU Leuven, Celestijnenlaan 200D, 3001, Leuven, Belgium\\
$^{3}$Universit\'e Grenoble Alpes, CNRS, IPAG, F-38000 Grenoble, France\\
$^{4}$Department of Physics and Space Science, Royal Military College of Canada, PO Box 17000, Stn Forces, Kingston, Ontario K7K 7B4, Canada
}  

\begin{document}

\date{Accepted 2018 April 27. Received 2018 April 27; in original form 2018 March 06}

\pagerange{\pageref{firstpage}--\pageref{lastpage}} \pubyear{2018}

\maketitle

\label{firstpage}

\begin{abstract}
Early-type magnetic stars are rarely found in close binary systems. No such objects were known in eclipsing binaries prior to this study. Here we investigated the eclipsing, spectroscopic double-lined binary \hd, which exhibits out-of-eclipse photometric variations suggestive of surface brightness inhomogeneities typical of early-type magnetic stars. Using a new set of high-resolution spectropolarimetric observations, we discovered a weak magnetic field on the primary and found intrinsic, element-dependent variability in its spectral lines. The magnetic field structure of the primary is dominated by a nearly axisymmetric dipolar component with a polar field strength $B_{\rm d}\approx600$~G and an inclination with respect to the rotation axis of $\beta_{\rm d}=13\degr$. A weaker quadrupolar component is also likely to be present. We combined the radial velocity measurements derived from our spectra with archival optical photometry to determine fundamental masses (3.16 and 1.75 $M_\odot$) and radii (2.78 and 1.39 $R_\odot$) with a 1--3\% precision. We also obtained a refined estimate of the effective temperatures (13000 and 9000~K) and studied chemical abundances for both components with the help of disentangled spectra. We demonstrate that the primary component of \hd\ is a typical late-B magnetic chemically peculiar star with a non-uniform surface chemical abundance distribution. It is not an HgMn-type star as suggested by recent studies. The secondary is a metallic-line star showing neither a strong, global magnetic field nor intrinsic spectral variability. Fundamental parameters provided by our work for this interesting system open unique possibilities for probing interior structure, studying atomic diffusion, and constraining binary star evolution.
\end{abstract}


\section{Introduction}
\label{intro}

The phenomenon of chemical peculiarity is widespread among main sequence B and A stars. Quiescence in the radiative envelopes of these objects facilitates chemical segregation by the competing effects of radiative pressure, gravitational settling and, occasionally, accretion of ISM material, giving rise to several distinct types of chemically peculiar (CP) stars (see reviews by \citealt{smith:1996}; \citealt{kurtz:2000}). A significant fraction of these objects \citep[up to $\sim$\,10\% of all OBA stars,][]{sikora:2018} possesses prominent, globally organised, kG-strength magnetic fields. These fields, thought to have a fossil origin \citep{braithwaite:2004,neiner:2015}, must have been acquired by stars early in their evolutionary history and remained essentially unchanged in the course of their main sequence life. 

The magnetic CP (Ap/Bp) stars are characterised by extreme chemical anomalies and exhibit conspicuous spectral and photometric rotational variability due to inhomogeneous chemical abundance distributions (spots of element over- and underabundance) on their surfaces. On the other hand, the so-called non-magnetic CP stars (cooler Am and hotter HgMn as well as PGa objects) show less extreme abundance peculiarities, little or no surface inhomogeneities and at least two order of magnitude weaker magnetic fields, which likely have a different physical origin than the magnetism of Ap/Bp stars. 

It is remarkable that the magnetic and non-magnetic CP stars have drastically different binary characteristics. Whereas the binary frequency of Am and HgMn stars is the same or higher than that of normal stars, Ap/Bp stars are almost entirely absent in close binary systems \citep{gerbaldi:1985,carrier:2002}. For example, among hundreds of known magnetic BA stars only about ten belong to spectroscopic binaries with $P_{\rm orb} < 20$~d \citep{landstreet:2017} and only one doubly-magnetic early-type binary, $\varepsilon$~Lup, has been found so far \citep{shultz:2015}. 

This unusually low binary incidence of hot magnetic stars suggests that stellar multiplicity is somehow linked with the absence of fossil magnetic fields. Several hypotheses explaining this anticorrelation are discussed in the literature. For instance, numerical simulations of the early massive-star formation stages indicate that the presence of a strong global primordial magnetic field inhibits protostellar cloud fragmentation, disfavouring formation of multiple systems \citep{commercon:2011}. On the other hand, it has been proposed that the field itself originates in the process of pre-main sequence binary star mergers \citep{schneider:2016}. In either case, short-period, early-type, magnetic binary systems are very unusual objects. Despite their low incidence, their sheer existence illuminates uncommon channels of the evolution of massive and intermediate-mass stars. These systems also represent unique astrophysical laboratories which enable us to derive useful constraints on the fundamental parameters of the binary companions. These constraints are particularly valuable considering the peculiar, non-standard structure of the chemically-stratified atmospheres and outer envelopes of magnetic CP stars.

\hd\ (V414~Pup, HIP\,39229), $V=8.8$, is listed as an A0 Si object in the catalogue of CP stars by \citet{renson:2009}. Based on the Hipparcos \citep{perryman:1997} and ASAS-3 \citep{pojmanski:2002} photometry, \citet{otero:2003} showed the presence of eclipses in the light curve of \hd\ accompanied by a smooth, stable, synchronous out-of-eclipse variability. This observation suggested that one of the binary components in this system exhibits an intrinsic rotational variability due to surface spots. Further photometric analysis of \hd\ by \citet{hummerich:2016} reinforced the conclusions by \citet{otero:2003} and confirmed that the primary component is a late-B chemically peculiar star with enhanced Si~{\sc ii} lines. This spectroscopic characteristic is a typical signature of the magnetic Bp stars, raising the intriguing possibility that the primary of \hd\ is a CP star with a global magnetic field.

Another photometric and spectroscopic study of \hd\ was carried out by \citet[][hereafter N17]{niemczura:2017}. These authors obtained  photometric time-series observations in several filters, determined atmospheric parameters of both components and presented detailed abundance analysis of the primary based on two spectroscopic observations. They argued that the primary of \hd\ is, in fact, not a magnetic Bp star but an object closely related to HgMn stars. The latter sub-group of late-B CP stars overlaps with the magnetic Bp stars in the H-R diagram but lacks global magnetic fields \citep{auriere:2010a,kochukhov:2013a}. Moreover, unlike Bp stars, HgMn stars are commonly found in close binaries \citep{smith:1996,scholler:2010,folsom:2013b}, including eclipsing systems \citep[e.g.][]{folsom:2010}. These stars exhibit moderate atmospheric abundance anomalies and weak surface inhomogeneities \citep{makaganiuk:2011,korhonen:2013}, typically detectable only with a high signal-to-noise ratio, high-resolution spectroscopy \citep[e.g.][]{adelman:2002,kochukhov:2005b,kochukhov:2011b} or with high-precision space-based photometry \citep{morel:2014,strassmeier:2017}.

\citet{paunzen:2018} combined the photometric observations of \hd\ by \citetalias{niemczura:2017} with 12 radial velocity measurements obtained from high- and medium-resolution spectra. Based on this dataset, they derived astrophysical parameters of the system and compared these results with the predictions of stellar evolutionary models.

In this paper we investigate the nature of \hd\ with the help of new high-resolution, time-series spectroscopic and spectropolarimetric observations. We report discovery of a global magnetic field and spectral variability of the primary component, demonstrating it to be a typical magnetic Bp star rather than a HgMn-related object as hypothesised by \citetalias{niemczura:2017}. Furthermore, we combine archival photometry with our new spectroscopic radial velocity (RV) measurements to derive precise fundamental masses and radii for both components of \hd. This makes the primary of this system the first magnetic Bp star for which such fundamental constraints have become available.

The rest of our paper is organised as follows. Sect.~\ref{obs} discusses the observational material used in our study. Sect.~\ref{analysis} presents analysis of these data, including investigation of the magnetic field, line profile variability, simultaneous RV and light curve binary star modelling, and assessment of the atmospheric parameters and abundances. The main conclusions of our study are summarised and discussed in Sect.~\ref{disc}.

\section{Observational data}
\label{obs}

\begin{table*}
\caption{Journal of spectropolarimetric observations of \hd. The columns indicate the date of observation, Heliocentric Julian Date at the middle of the observing sequence, orbital phase, median S/N ratio of the spectra in the 5000--6000~\AA\ wavelength interval, S/N of LSD profiles, radial velocities of the primary and secondary, mean longitudinal magnetic field of the primary, false alarm probability for the Stokes $V$ signature of the primary, and the corresponding field detection significance (ND=no detection, DD=definite detection).}
\centering
\begin{tabular}{l l l l r r r r l l}
\hline
\hline
UT date & HJD & Phase & S/N$_{\rm obs}$ & S/N$_{\rm LSD}$ & $V_A$ (\kms) & $V_B$ (\kms) & \bz\ (G) & FAP & Detection \\
\hline
2016-12-15 & 2457737.9971 & 0.852 & 253 & 10661 & $ 59.8$ & $-113.5$ & $ 32\pm26$ & 0.0                 &  DD \\
2016-12-21 & 2457744.0766 & 0.132 & 241 & 10364 & $-57.2$ & $  99.9$ & $ 54\pm27$ & $5.1\times10^{-1}$  &  ND \\
2016-12-22 & 2457744.9862 & 0.323 & 261 & 11278 & $-71.3$ & $ 120.0$ & $  1\pm26$ & $7.4\times10^{-1}$  &  ND \\
2017-01-08 & 2457761.9942 & 0.905 & 225 &  9741 & $ 41.6$ & $ -79.7$ & $ 10\pm28$ & 0.0                 &  DD \\
2017-01-09 & 2457763.0279 & 0.122 & 257 & 11037 & $-54.1$ & $  94.9$ & $ 23\pm25$ & $1.5\times10^{-1}$  &  ND \\
2017-01-10 & 2457764.0415 & 0.336 & 256 & 11176 & $-68.4$ & $ 116.5$ & $ -5\pm26$ & $1.1\times10^{-1}$  &  ND \\
2017-01-11 & 2457765.0367 & 0.545 & 233 &  9934 & $ 19.4$ & $ -40.6$ & $ 10\pm29$ & $7.9\times10^{-1}$  &  ND \\
2017-01-12 & 2457766.0215 & 0.753 & 250 & 10884 & $ 75.3$ & $-138.7$ & $-39\pm26$ & $2.9\times10^{-12}$ &  DD \\
2017-01-13 & 2457767.0361 & 0.966 & 270 & 11782 & $ 14.8$ & $ -32.1$ & $ 98\pm23$ & 0.0                 &  DD \\
2017-01-14 & 2457767.9954 & 0.168 & 247 & 10566 & $-66.9$ & $ 115.9$ & $ 24\pm27$ & $8.0\times10^{-1}$  &  ND \\
2017-01-15 & 2457769.0496 & 0.390 & 241 & 10321 & $-52.7$ & $  87.0$ & $ 37\pm29$ & $1.3\times10^{-3}$  &  ND \\
2017-01-16 & 2457770.0308 & 0.597 & 205 &  8648 & $ 41.3$ & $ -80.4$ & $-33\pm32$ & $9.8\times10^{-1}$  &  ND \\
2017-01-17 & 2457771.0382 & 0.809 & 372 & 15559 & $ 69.7$ & $-132.0$ & $-42\pm18$ & 0.0                 &  DD \\
2017-01-17 & 2457771.0790 & 0.817 & 314 & 13033 & $ 68.1$ & $-130.1$ & $-31\pm21$ & 0.0                 &  DD \\
\hline
\end{tabular}
\label{tbl:obs}
\end{table*}

We observed \hd\ in December 2016 and January 2017 with the ESPaDOnS spectropolarimeter at the Canada-France-Hawaii Telescope (CFHT). These observations were carried out in the framework of the BinaMIcS CFHT large program \citep{alecian:2015a}. Fourteen circular polarisation observations of \hd\ were acquired on thirteen individual nights. Each polarimetric measurement comprised four subexposures obtained with two different polarimeter configurations in order to exchange the path of the orthogonal polarisation beams through the system and their position on the detector. Twelve observations were obtained with a 300~s subexposure time (yielding 1200~s total exposure time); two measurements were obtained with a 840~s subexposure time (3360~s total exposure time). 

All spectra were reduced with the Libre-ESpRIT software \citep{donati:1997} running at the telescope. The intensity (Stokes $I$) spectra were obtained by adding together orthogonal polarisation beams from all four subexposures. The circular polarisation (Stokes $V$) spectra were derived using the ratio method \citep{bagnulo:2009} to minimise systematic errors due to instrumental artefacts. The same four subexposures were also used to obtain a diagnostic null spectrum in order to assess residual systematic errors. Continuum normalisation was performed by iteratively fitting a smooth function to the upper envelope of the merged stellar spectrum \citep{rosen:2018}, excluding telluric and hydrogen line regions. The resulting spectra have a resolving power of 65000 and provide a nearly complete coverage of the 3700--10480~\AA\ wavelength range. The signal-to-noise (S/N) ratio varies from 200 to 370 and has a median value of 250. 

The complete log of our spectropolarimetric observations of \hd\ is presented in Table~\ref{tbl:obs}. Columns 1 to 4 give the UT date of observation, the heliocentric Julian date corresponding to the middle of observation, the orbital phase, and the median S/N ratio per 1.8~\kms\ velocity bin in the 5000--6000~\AA\ wavelength region. The orbital phase was calculated according to the ephemeris $\mathrm{HJD}=2452167.8724+E\times4.7492148$ determined in Sect.~\ref{binary}. The same ephemeris gives the rotational phase of the primary (the hotter, more massive and more luminous component) since, as confirmed later in the paper, its rotation is synchronised with the orbital motion. 

In addition to the ESPaDOnS high-resolution spectropolarimetric observations we made use of the $B$, $V$, and $I_{\rm c}$ time-series photometry published by \citetalias{niemczura:2017}. We refer the reader to that paper for details of the acquisition and reduction of the photometric data. In addition, we made use of the $V$-band photometric observations from the ASAS-3 project \citep{pojmanski:2002}.

\section{Analysis}
\label{analysis}

\subsection{Least-squares deconvolved profiles}

\subsubsection{Calculation of LSD profiles}
\label{calclsd}

No circular polarisation signal is evident in individual metal lines in any of our Stokes $V$ spectra of \hd. To enhance the S/N of the polarisation profiles and to derive average Stokes $I$ spectra amenable to straightforward RV measurements, we employed the least-squares deconvolution (LSD) code developed by \citet{kochukhov:2010a}. This software performs an intelligent co-addition of information from all usable metal lines following the principles described by \citet{donati:1997}. In this widely used method, each intensity or polarisation spectral feature is assumed to be a shifted and scaled copy of a mean profile. It is also assumed that contributions of overlapping lines add up linearly. The line scaling factors, or weights, are given by the theoretical line depth for Stokes $I$ and by the product of the line depth, central wavelength and the effective Land\'e factor for Stokes $V$. Provided a list of spectral line positions and weights (a line mask) corresponding to a specified set of stellar atmospheric parameters, the LSD algorithm allows one to quickly derive a high S/N ratio mean profile for a given observed spectrum and its error bars. \citet{kochukhov:2010a} showed that approximations inherent to the LSD technique are appropriate for a Stokes $I$ and $V$ profile analysis if the local magnetic field strength does not exceed $\sim$\,2~kG.

We used the VALD database \citep{ryabchikova:2015} to retrieve a line list for $T_{\rm eff}=12500$~K, $\log g=4.0$, and element abundances of \hd\ A as reported by \citetalias{niemczura:2017}. The lines with intrinsic depths less than 10\% as well as lines located in the broad hydrogen line wings and in regions contaminated by the telluric absorption were removed. The final LSD line mask included 2359 lines, dominated by Fe~{\sc ii}, with the mean wavelength $\lambda_0=5160$~\AA\ and mean effective Land\'e factor $z_0=1.18$. The same $\lambda_0$ and $z_0$ were used for the normalisation of the LSD profiles.

The LSD procedure was applied to all Stokes $I$, $V$ and diagnostic null spectra. The mean profiles were calculated in the $\pm350$~\kms\ velocity range, using a 2~\kms\ velocity bin. The resulting S/N ratio of the LSD profiles, reported in the 5th column of Table~\ref{tbl:obs}, indicates a factor of $\approx$\,40 gain with respect to the S/N ratio of the original spectra.

\begin{figure}
\centering
\figps{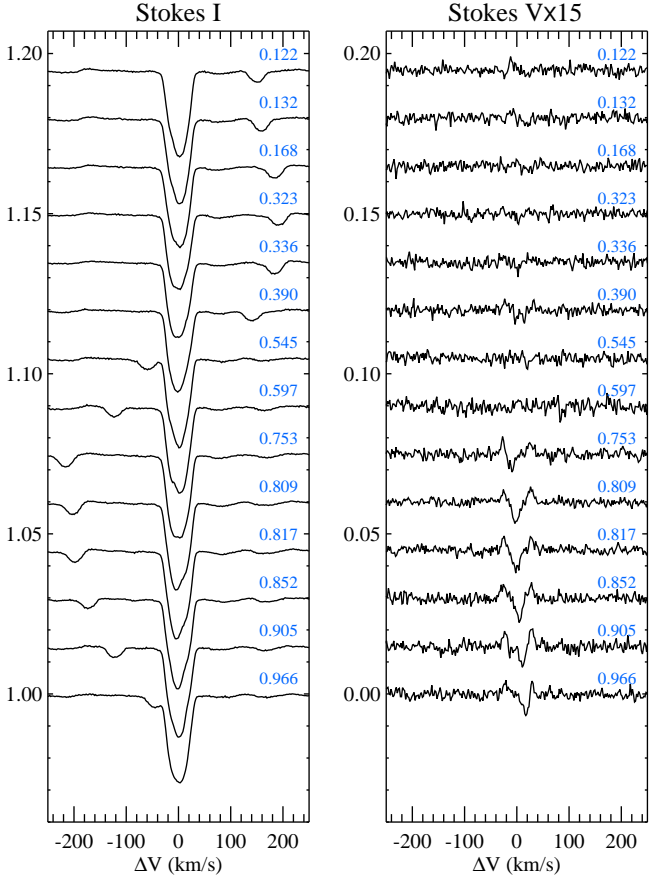}
\caption{Least-squares deconvolved Stokes $I$ (left panel) and Stokes $V$ (right panel) profiles of \hd. The spectra are shifted to the laboratory frame of the primary component and offset vertically according to the orbital phase indicated next to each profile. The Stokes $V$ profiles are amplified by a factor of 15.}
\label{fig:lsd}
\end{figure}

\subsubsection{Radial velocity measurements}
\label{rvlsd}

The Stokes $I$ LSD profiles obtained using the line-addition procedure outlined above are displayed in the left panel of Fig.~\ref{fig:lsd}. The prominent, slightly variable, absorption signature of the primary component is accompanied by a much weaker contribution of the secondary. The latter appears to be constant in shape and is found to be red-shifted with respect to the primary's profile for phases 0.122--0.390 and blue-shifted for phases 0.545--0.966, in full agreement with the relative radial velocities expected for a circular orbital motion.

A simple mean profile disentangling (component separation) procedure adapted from the method used by \citet{folsom:2010,folsom:2013a} was applied to measure RVs of the \hd\ components from the Stokes $I$ LSD profiles. This version of the disentangling algorithm determines individual, time-dependent radial velocities and mean component profiles under the assumption that the latter are constant in time. This approximation is not strictly correct for \hd\ A. Nevertheless, in our experience this method yields considerably more robust and accurate radial velocities when applied to stars with spectra moderately distorted by surface spots compared to measurements using the centre-of-gravity method or analytical function fitting \citep{rosen:2018}.

The radial velocities of \hd\ A and B determined with the help of the LSD profile disentangling technique are given in columns 6 and 7 of Table~\ref{tbl:obs}. No realistic error estimate is possible with our disentangling algorithm. But subsequent fitting of the orbital solution to these RV data suggests typical error of 0.7~\kms\ for the primary and 1.3~\kms\ for the secondary.

\subsubsection{Polarisation signature detection}
\label{fap}

Using the radial velocities of the binary components derived from the Stokes $I$ LSD profiles, we examined the Stokes $V$ LSD spectra in the laboratory frame of the primary component (see right panel of Fig.~\ref{fig:lsd}). This analysis reveals clear polarisation signatures at the radial velocity of the primary between orbital phases 0.753 and 0.966. These signatures have no counterparts in the LSD profiles obtained from the diagnostic null spectrum. At the same time, no evidence of a magnetic signature is found at the position of the secondary component. Consequently, these results represent an unambiguous detection of the global magnetic field in \hd\ A and no detection of magnetic field in \hd\ B. 

Application of the formal false alarm $\chi^2$ probability (FAP) analysis \citep{donati:1992,donati:1997} to the $\pm40$~\kms\ velocity interval around the position of the primary yields a definite magnetic field detection (FAP\,$<$\,$10^{-5}$) for six Stokes $V$ profiles and a non-detection (FAP\,$>$\,$10^{-3}$) for the remaining eight profiles. Individual FAP values are given in the 9th column of Table~\ref{tbl:obs}. Identical analysis applied to the null LSD profiles yields only non-detections, with FAP\,$\ge$\,$8.4\times10^{-1}$.

\subsection{Magnetic field strength and topology}
\label{mag}

\subsubsection{Longitudinal magnetic field}

\begin{figure}
\centering
\figps{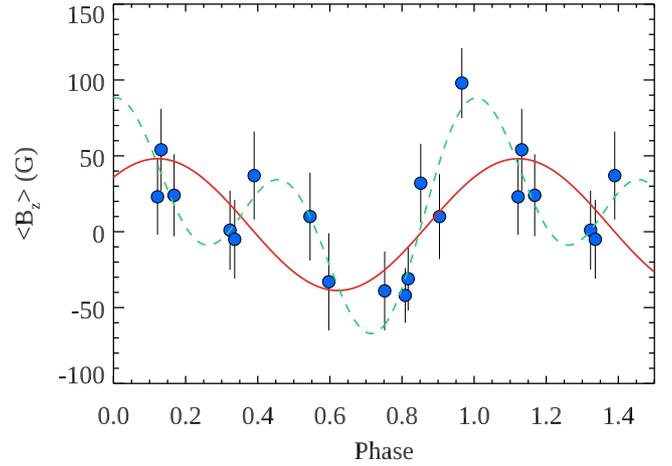}
\caption{Mean longitudinal magnetic field of the primary component of \hd\ as a function of rotational phase. The curves illustrate the single-wave (red solid line) and double-wave (green dashed line) Fourier fits.}
\label{fig:bz}
\end{figure}

To obtain quantitative information about the strength and topology of the global magnetic field of \hd\ A we measured the mean longitudinal magnetic field, \bz, from the Stokes $V$ LSD profiles. The Stokes $I$ signature of the secondary was subtracted prior to these measurements. The mean longitudinal field, which provides a measure of the disk-averaged line-of-sight magnetic field component, was obtained by computing the first moment of the Stokes $V$ profile and normalising it by the equivalent width of the Stokes $I$ profile \citep[e.g.][]{kochukhov:2010a}
\begin{equation}
\langle B_{\rm z} \rangle  = -7.145 \times 10^6 \dfrac{\int (v-v_0) V \mathrm{d}v}{\lambda_0 z_0 \int (1-I) \mathrm{d}v},
\label{eq:bz}
\end{equation}
where the velocity coordinate $v$ and the centre-of-gravity velocity of the $I$ profile, $v_0$, are measured in \kms\ and the result is in gauss. Owing to the normalisation of the Stokes $V$ moment by the equivalent width of the intensity profile, \bz\ calculated with this formula is unaffected by continuum contribution of a binary companion or any other continuum light source.

\begin{figure*}
\centering
\fifps{16cm}{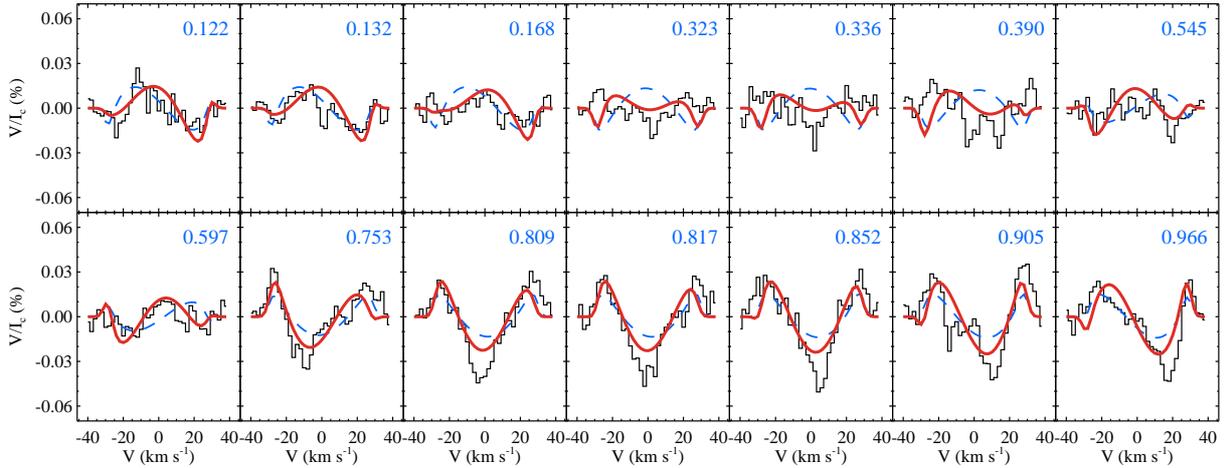}
\caption{Comparison of the observed (black histogram) LSD Stokes $V$ profiles of \hd\ with the model profiles for the dipole (blue dashed line) and dipole+quadrupole (red solid line) magnetic field geometries discussed in the text. Rotational phases are indicated in the upper right corner of each sub-panel.}
\label{fig:gstokes_prf}
\end{figure*}

We evaluated the integrals in Eq.~(\ref{eq:bz}) with the help of the trapezoidal integration scheme, within $\pm40$~\kms\ of the centre-of-gravity of the primary line. The \bz\ error was deduced following the standard error propagation rules. The resulting longitudinal field measurements are reported in the 8th column of Table~\ref{tbl:obs}. The measured \bz\ values span from about $-40$~G to 100~G. The median error is 26~G, which is comparable to \bz\ values themselves. In fact, only one longitudinal field measurement at phase 0.966 corresponds to a $>$3$\sigma$ \bz\ detection. This is because several observed Stokes $V$ profiles show nearly symmetric polarisation signatures. In this case, longitudinal field measurements do not allow one to confidently ascertain the presence of a magnetic field (notwithstanding that the Stokes $V$ profiles firmly indicate that a field is present).

The same longitudinal field measurement procedure was applied to the diagnostic null LSD profiles. No formal \bz\ detections were obtained, with the most significant \bz\,(N) value being about 1.6$\sigma$. The standard deviation of the fourteen \bz\,(N) values is 21~G, in good agreement with the typical formal error of the Stokes $V$ \bz\ measurements inferred above.

The mean longitudinal magnetic field of \hd\ A is illustrated as a function of phase in Fig.~\ref{fig:bz}. Despite the lack of individual \bz\ detections, the longitudinal field appears to follow a coherent phase variation. In particular, the minimum around phase 0.7 is well-defined. By fitting this variation with a sinusoid function (solid curve in Fig.~\ref{fig:bz}) we found a constant term $B_0=4.7$~G and a semi-amplitude of $B_1=43$~G. This fit may be considered unsatisfactory given its reduced chi-square ($\chi^2_\nu$) of 1.9. Adding the first harmonic results in a much better description of the observed \bz\ variation (dashed line in Fig.~\ref{fig:bz}). This fit yields $\chi^2_\nu$ of 0.7, probably indicating overfitting of the data.

To summarise, this analysis of the longitudinal magnetic field of \hd\ A does not offer much insight into the field characteristics apart from suggesting that the polar field intensity is on the order of a few hundred G and that the field topology is likely not purely dipolar.

\subsubsection{Stokes $V$ profile modelling}

Direct modelling of the LSD Stokes $V$ profile signatures offers an alternative approach to obtaining quantitative characteristics of the global magnetic field of the primary component of \hd. Such an analysis is also necessary to address the puzzling phase dependence of the observed Stokes $V$ profile amplitude, with only a handful of observations in a limited phase interval showing a definite polarisation signature.

The phase coverage and quality of the available LSD profiles is clearly insufficient for a detailed Zeeman Doppler imaging study \citep[e.g.][]{kochukhov:2014,kochukhov:2017a}. Instead, a simpler multipolar fit to the circular polarisation data \citep[e.g.][]{alecian:2008a,alecian:2016} can be used to get an idea about the field strength and geometry. In this study we performed such modelling with the help of the {\sc GStokes} code\footnote{\url{http://www.astro.uu.se/~oleg/gstokes.html}}. This tool, written in IDL and equipped with a user-friendly graphical front-end, enables forward calculation of the disk-integrated Stokes parameter profiles as well as magnetic inversions under several widely used simplifying approximations of the polarised line formation. The code implements the Unno-Rachkovsky analytical solution of the polarised radiative transfer equation \citep[e.g.][]{polarization:2004} and the weak-field approximation with the Gaussian local profiles \citep[e.g.][]{petit:2012}. The magnetic field geometry is described with one of the common low-order multipolar field parameterisations: a centred or offset dipole \citep{achilleos:1989}, a superposition of the aligned, axisymmetric dipole, quadrupole and octupole components \citep{landstreet:1988}, and a dipole plus general, non-axisymmetric quadrupolar field \citep{bagnulo:1996}. Various stellar (line depth, \vs, radial velocity, inclination) and magnetic field parameters can be adjusted for a given set of observed LSD profiles using a powerful non-linear least-squares optimisation algorithm \citep{markwardt:2009}.

The present analysis of \hd\ A was carried out using Gaussian local Stokes $I$ profiles with $FWHM=5$~\kms\ and calculating the Stokes $V$ profiles under the weak-field approximation with $\lambda=5160$~\AA\ and $z=1.18$. The linear limb-darkening with a coefficient of 0.37 was also adopted. The stellar rotational axis was assumed to be aligned with the orbital axis. Consequently, the inclination angle of the former was fixed to the value of 86.14\degr\ according to the binary system modelling in Sect.~\ref{binary}. In the first step of the analysis the local equivalent width of the intensity profiles, assumed to be constant over the stellar surface, and \vs\ were adjusted to reproduce the LSD Stokes $I$ spectra. Then, the magnetic field topology was derived from the Stokes $V$ spectra.

Initially, we attempted to reproduced the observed Stokes $V$ profiles with a centred dipolar field. This analysis yields a polar field strength of $B_{\rm d}=472\pm111$~G and magnetic obliquity of $\beta_{\rm d}=18\pm4$\degr. However, the corresponding fit (dashed line in Fig.~\ref{fig:gstokes_prf}) to the observed LSD profiles is not entirely satisfactory, which is attested to by a fairly high value of the reduced chi-square, $\chi^2_\nu=2.0$, and inability of the model profiles to reproduce the full extent of the rotational modulation of the Stokes $V$ signature. Adding a non-axisymmetric quadrupolar component improves the fit significantly. The $\chi^2_\nu$ is reduced to 1.3 and theoretical profiles achieve a reasonable description of the observations (see solid curve in Fig.~\ref{fig:gstokes_prf}), even including rotational phases for which no formal detection of the polarisation signatures was obtained in Sect.~\ref{fap}. In this case, the best-fitting dipolar field parameters are $B_{\rm d}=625\pm172$~G and $\beta_{\rm d}=12\pm3$\degr. The strength of the quadrupolar component is found to be $B_{\rm q}=214\pm43$~G.

\begin{figure}
\centering
\firps{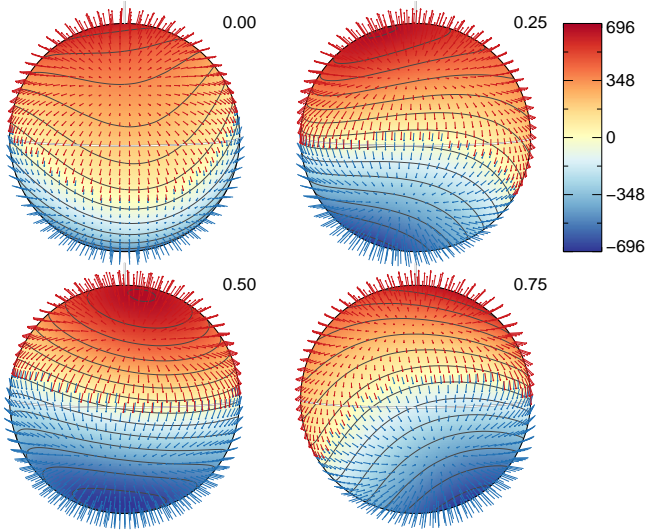}
\caption{Magnetic field topology of the primary component of \hd. Distribution of the radial magnetic field component (colour plot) and the field vector orientation (red and blue vectors) are shown at four rotation phases. The field strength is given in gauss. The star is shown at the inclination angle of $86\degr$.}
\label{fig:gstokes_fld}
\end{figure}

The final global magnetic field geometry derived for the primary component of \hd\ is illustrated in Fig.~\ref{fig:gstokes_fld}. The field structure is nearly axisymmetric and dipole-dominated. The field intensity is relatively low, with the maximum local field strength of about 700~G and the surface-averaged field modulus of 443~G.

\subsection{Binary star model}
\label{binary}

We analysed the observed light curves and radial velocity variation of \hd\ with a physical, Roche geometry model of the binary system implemented in the {\sc PHOEBE} code\footnote{\url{http://phoebe-project.org/}} \citep{prsa:2005,prsa:2016}. 
Details of our application of the {\sc PHOEBE} code and the accompanying Bayesian error analysis are described in Appendix.~\ref{phoebe}.

\begin{table}
\caption{\label{parameters_table}Free parameters of the {\sc PHOEBE} model and the corresponding priors used in the Bayesian optimisation and error analysis. For each parameter we list the units, when applicable, the priors, and the best-fitting values together with the 95\% confidence intervals.}
\centering
{\small
\begin{tabular}{lcc}
\hline\hline
Parameter&Prior&Best fit\\
\hline
\rule{0pt}{2.5ex}$L^{(1)}_{B}/L^{(1+2)}_{B}$ [\%] & $\left(50,100\%\right)$ & $91.4817\substack{+0.0035 \\ -0.0035}$ \\
\rule{0pt}{2.5ex}$L^{(1)}_{V}/L^{(1+2)}_{V}$ [\%] & $\left(50,100\%\right)$ & $90.5396\substack{+0.0027 \\ -0.0032}$ \\
\rule{0pt}{2.5ex}$L^{(1)}_{I_{\rm c}}/L^{(1+2)}_{I_{\rm c}}$ [\%] & $\left(50,100\%\right)$ & $89.1629\substack{+0.0065 \\ -0.0065}$ \\
\rule{0pt}{2.5ex}$L^{(1)}_{ASAS-V}/L^{(1+2)}_{ASAS-V}$ [\%] & $\left(50,100\%\right)$ & $90.7\substack{+2.1 \\ -2.8}$ \\
\rule{0pt}{2.5ex}$i \,\mathrm{[\degr]}$  & $\left(75,90\right)$ & $86.141\substack{+0.087 \\ -0.073}$ \\
\rule{0pt}{2.5ex}$M_2/M_1$  & $\left(0,1\right)$ & $0.5549\substack{+0.0063 \\ -0.0068}$ \\
\rule{0pt}{2.5ex}$a\,[R_{\odot}]$  & $\left(15,25\right)$ & $20.203\substack{+0.098 \\ -0.096}$ \\
\rule{0pt}{2.5ex}$\mathrm{\gamma}$ [\kms] & $\left(-10,10\right)$ & $-2.14\substack{+0.37 \\ -0.36}$ \\
\rule{0pt}{2.5ex}$\mathrm{HJD_{0}}$\,[d]  & $\left(2452167.87\pm0.2\right)$ & $2452167.8724\substack{+0.0017 \\ -0.0017}$ \\
\rule{0pt}{2.5ex}$P_{\rm orb}$\,[d] & $\left(4.7,4.8\right)$ & $4.7492148\substack{+0.0000015 \\ -0.0000015}$ \\
\rule{0pt}{2.5ex}$T^{(1)}_{\rm eff}$\,[K]  & $\left(12500,500\right)$ & $12100\substack{+1000 \\ -900}$ \\
\rule{0pt}{2.5ex}$\Omega_{1}$  & $\left(4,15\right)$ & $7.831\substack{+0.048 \\ -0.047}$ \\
\rule{0pt}{2.5ex}$T^{(2)}_{\rm eff}$\,[K]  & $\left(8000,500\right)$ & $8100\substack{+0900 \\ -700}$ \\
\rule{0pt}{2.5ex}$\Omega_{2}$  & $\left(4,15\right)$ & $9.29\substack{+0.11 \\ -0.11}$ \\
\hline
\end{tabular}
}
\end{table}

Our {\sc PHOEBE} analysis was performed using the four archival light curves and fourteen RV measurements obtained in our study (see Table~\ref{tbl:obs}). The first three light curves are taken directly from \citetalias{niemczura:2017}, whereas the fourth one was obtained from the ASAS-3 $V$-filter database \citep{pojmanski:2002}. As discussed earlier, the photometry reveals out-of-eclipse modulation, which is thought to be produced by spots on the surface of the primary star. Non-uniform brightness distributions in Ap/Bp stars are caused by surface chemical inhomogeneities induced by strong magnetic fields. These structures are thought to be stable for at least several decades. Although {\sc PHOEBE} includes a spot modelling option, it does not incorporate a physical model of the metallicity-dependent flux redistribution required for a quantitative interpretation of the photometric variations of Ap/Bp stars in different bands \citep[e.g.][]{shulyak:2010b,krticka:2012}. For this reason, and to avoid introducing additional free (and degenerate) spot parameters, we chose to remove the spot-induced variations from all the photometric data prior to {\sc PHOEBE} modelling. This was accomplished by prewhitening the photometric data with second-order Fourier functions fitted to the out-of-eclipse regions of each light curve.

\begin{figure}
\centering
\figps{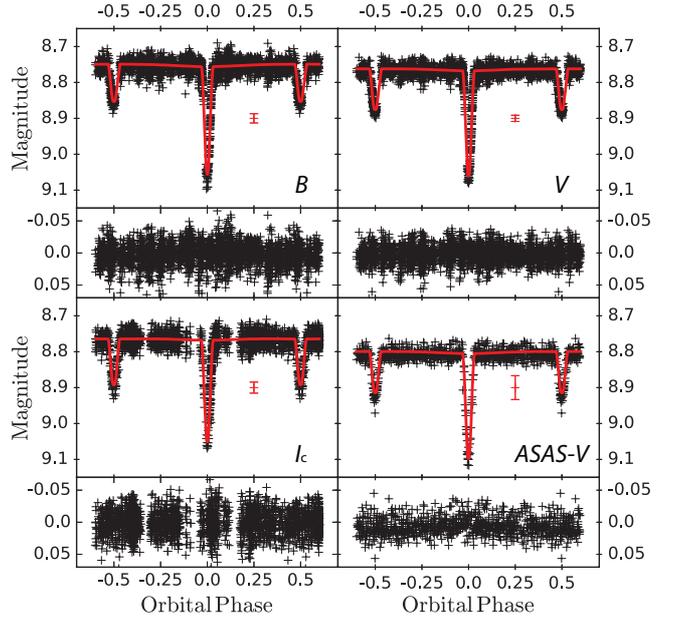}
\caption{Comparison of the photometric measurements of \hd\ in different bands (black crosses) with the {\sc PHOEBE} light curve models (red solid curves) corresponding to the median parameter values reported in Table \ref{parameters_table}. The panels show Johnson $B$ (\textit{upper left}), $V$ (\textit{upper right}), and Cousins $I_c$ (\textit{lower left}) with observations from \citetalias{niemczura:2017} and Johnson $V$ (\textit{lower right}) with data from the ASAS-3 database \citep{pojmanski:2002}. The average error bar is shown in red. Residuals are displayed below each light curve.}
\label{fig:lc_model}
\end{figure}

{\sc PHOEBE} addresses limb darkening at both the bolometric and individual filter level. The code allows for interpolation of limb darkening coefficients, computed for model atmosphere grids according to several laws, of which we chose the square-root law. We adopted the gravity-darkening coefficients per filter from \citet{claret:2011}, according to the effective temperatures and surface gravities reported by \citetalias{niemczura:2017}. We did not include any third light and calculate un-binned models. The prior parameter distributions employed for the Markov Chain Monte Carlo analysis described in Appendix~\ref{phoebe} are indicated in Table \ref{parameters_table}. Uniform priors in the specified ranges were adopted for all parameters except $T_{\rm eff}$, for which we used Gaussian priors according to the \citetalias{niemczura:2017} results. (Our own spectroscopic analysis presented below largely confirms these temperatures.) The second-order Fourier term removed to account for the photometric spot modulation also effectively removes all signal associated with the rotation of either component due to the tidal synchronisation of the system. Therefore, any modulation caused by varying the albedo of either component is also removed. For this reason, we choose to fix the albedo of both components to unity.

\begin{figure}
\centering
\figps{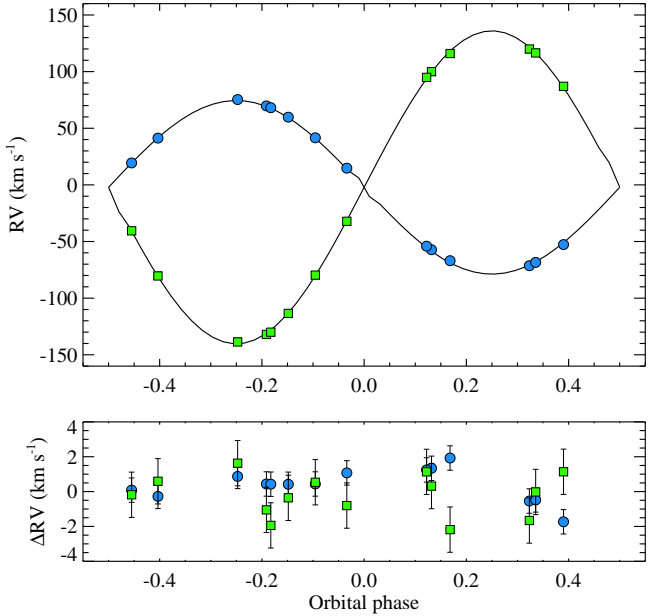}
\caption{Comparison of the radial velocity measurements for the primary (blue circles) and secondary (green squares) component of \hd\ with the {\sc PHOEBE} model radial velocity curves (solid lines). The bottom panel shows residuals after subtracting the best-fitting model from the data.}
\label{fig:rv_model}
\end{figure}

The final {\sc PHOEBE} models and residuals for each observed dataset (light curves and RVs) are shown Figs.~\ref{fig:lc_model} and \ref{fig:rv_model}. The final model parameters as extracted from the posterior distributions (shown and discussed in Appendix~\ref{phoebe}) are listed in Table~\ref{parameters_table}. The derived fundamental stellar parameters are given in Table~\ref{derived_table}. All models show a good fit to observations with no structure to the residuals. We note that the distortions of the simulated RV curves close to phase 0.0 for the primary and phase 0.5 for the secondary seen in Fig.~\ref{fig:rv_model} are real. These features represent manifestation of the Rossiter-Mclaughlin effect \citep{rossiter:1924,mclaughlin:1924}.

\begin{table}
\caption{\label{derived_table}Derived physical parameters for the primary and secondary component.}
\centering
\begin{tabular}{lcc}
\hline\hline
Parameter & Primary & Secondary\\
\hline
\rule{0pt}{2.5ex}$M \, [M_{\odot}]$ & $3.155\substack{+0.060 \\ -0.057}$  & $1.751\substack{+0.038 \\ -0.039}$ \\
\rule{0pt}{2.5ex}$R \, [R_{\odot}]$ & $2.781\substack{+0.034 \\ -0.034}$  & $1.390\substack{+0.042 \\ -0.042}$ \\
\rule{0pt}{2.5ex}$\log g \,\mathrm{[dex]}$     & $4.049\substack{+0.011 \\ -0.010}$ & $4.395\substack{+0.021 \\ -0.021}$ \\
\hline
\end{tabular}
\end{table}

\subsection{Atmospheric parameters and abundances}
\label{abn}

\subsubsection{Luminosity ratio and hydrogen line profiles}
\label{hlines}

We determined atmospheric parameters of the \hd\ components taking advantage of their accurately determined masses, radii, and relative luminosities. According to the results of the previous Section, fundamental surface gravities are $\log g_1 =4.05$ and $\log g_2=4.40$ for the primary and secondary, respectively. Keeping these $\log g$ values fixed, we calculated a grid of {\sc LLmodels} \citep{shulyak:2004} atmospheric models using abundances of \hd\ A reported by \citetalias{niemczura:2017} and assuming the solar photospheric chemical composition \citep{asplund:2009} for the secondary. The effective temperatures were varied within a 12000--13500~K interval for the primary and a 8000--9500~K range for the secondary, respectively. The theoretical spectral energy distributions of both components were convolved with the response curves of the Johnson $B$, $V$ and Cousins $I_c$ filters \citep{mann:2015} and compared to the observed luminosity ratios in these three bands using $R_1/R_2=2.781/1.390=2.000$. This analysis allowed us to establish that the effective temperature of the secondary star given by $T_{\rm eff}^{(2)}=3507+0.419\times T_{\rm eff}^{(1)}$ provides a consistency with the observed luminosity ratios over the entire considered range of $T_{\rm eff}^{(1)}$.

We then refined the effective temperatures using the observed composite hydrogen line profiles. Theoretical spectra around the hydrogen Balmer lines were calculated with the {\sc Synth3} code \citep{kochukhov:2007d} for both components. These calculations were then combined using appropriate radial velocity shifts, continuum fluxes (also calculated with {\sc Synth3}), and the observed ratio of radii. The resulting composite synthetic spectra around the H$\alpha$, H$\beta$, and H$\gamma$ lines were compared with the observed spectra at phases 0.323 and 0.753. These observations were chosen because they correspond to the largest velocity separation between the components and hence exhibit the most clear asymmetry of the hydrogen line shapes due to contribution of the secondary. 

\begin{figure}
\centering
\fifps{7.5cm}{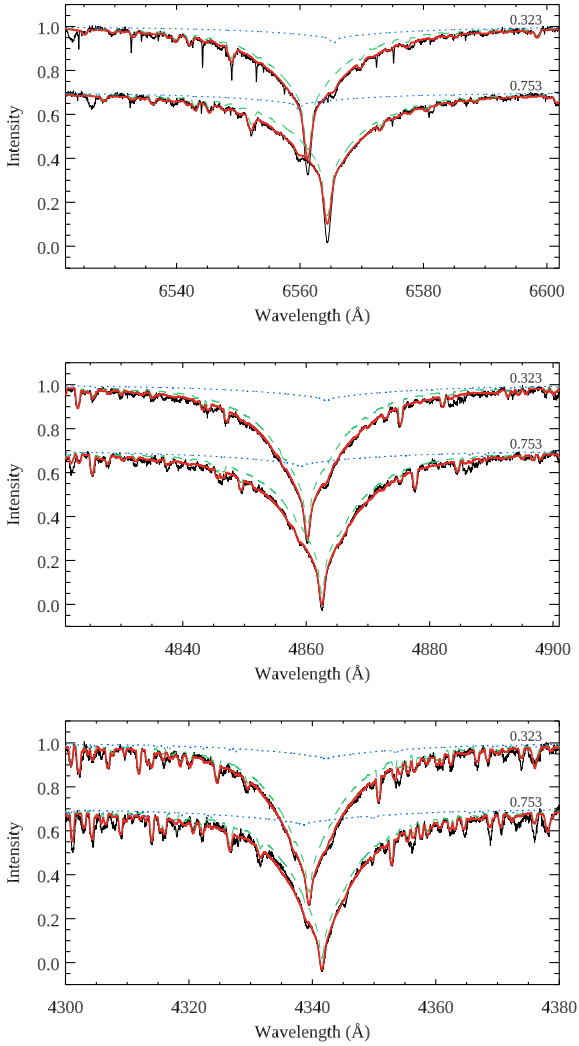}
\caption{The observed profiles of the H$\alpha$, H$\beta$, and H$\gamma$ hydrogen Balmer lines (thin solid black line) at the phases of maximum relative velocity shift between the binary components compared to the composite synthetic spectra (thick solid red line) computed for $T_{\rm eff}^{(1)}$\,=\,13000~K and $T_{\rm eff}^{(2)}$\,=\,9000~K. The contributions of the primary and secondary are shown with the green dashed and blue dotted lines, respectively. The profiles for phase 0.753 are offset vertically by $-0.15$.}
\label{fig:hlines}
\end{figure}

We found that, taking into account the previous constraint on the relative $T_{\rm eff}$ values, a good agreement with observations (see Fig.~\ref{fig:hlines}) is achieved for $T_{\rm eff}^{(1)}$\,=\,12700--13300~K and $T_{\rm eff}^{(2)}$\,=\,8800--9100~K. We therefore adopted $T_{\rm eff}^{(1)}$\,=\,$13000\pm300$~K and $T_{\rm eff}^{(2)}$\,=\,$9000\pm150$~K for the subsequent analysis. This determination of the effective temperature of the primary is broadly consistent with the 12050--12750~K range obtained by \citetalias{niemczura:2017} by applying different photometric calibrations to the unresolved Str\"omgren and Geneva photometric measurements of \hd. On the other hand, our $T_{\rm eff}^{(2)}$\,=\,9000~K is higher than $T_{\rm eff}^{(2)}$\,$\sim$\,8000~K suggested by these authors.

\subsubsection{Spectral disentangling}

\begin{figure}
\centering
\figps{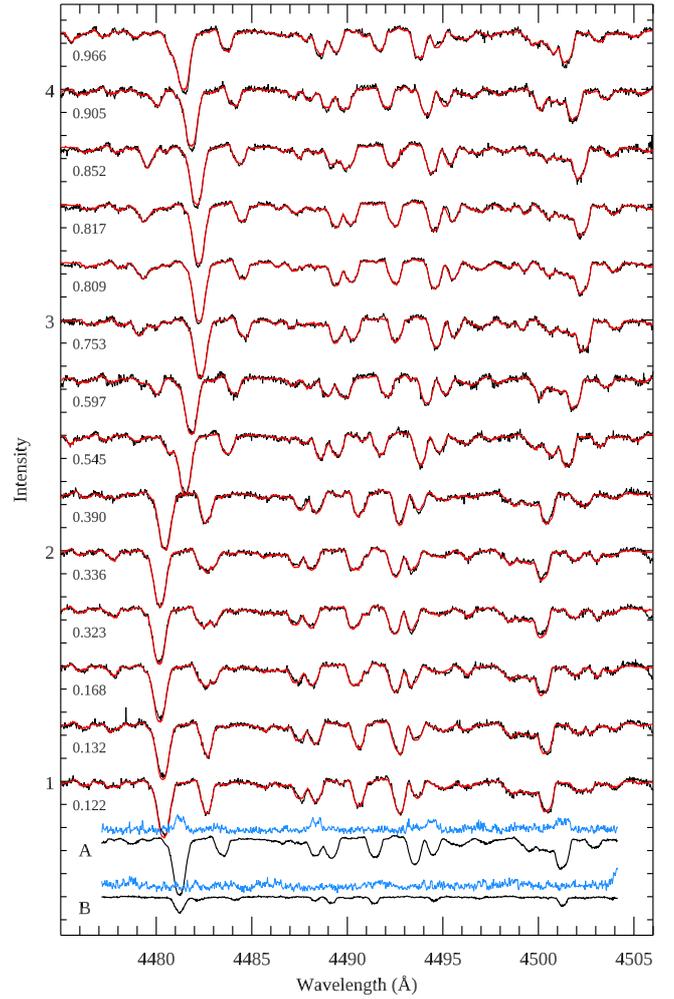}
\caption{Illustration of the spectral disentangling procedure in the 4475--4505~\AA\ wavelength region. The observed spectra are shown with the thin black lines. The thick red lines corresponds to the binary model spectra. The profiles are offset vertically and arranged according to the orbital phase indicated to the left. The final disentangled spectra (thick black curves) are shown at the bottom of the plot together with the shifted and scaled standard deviation profiles (light blue curves).}
\label{fig:sbfit}
\end{figure}

\begin{figure}
\centering
\figps{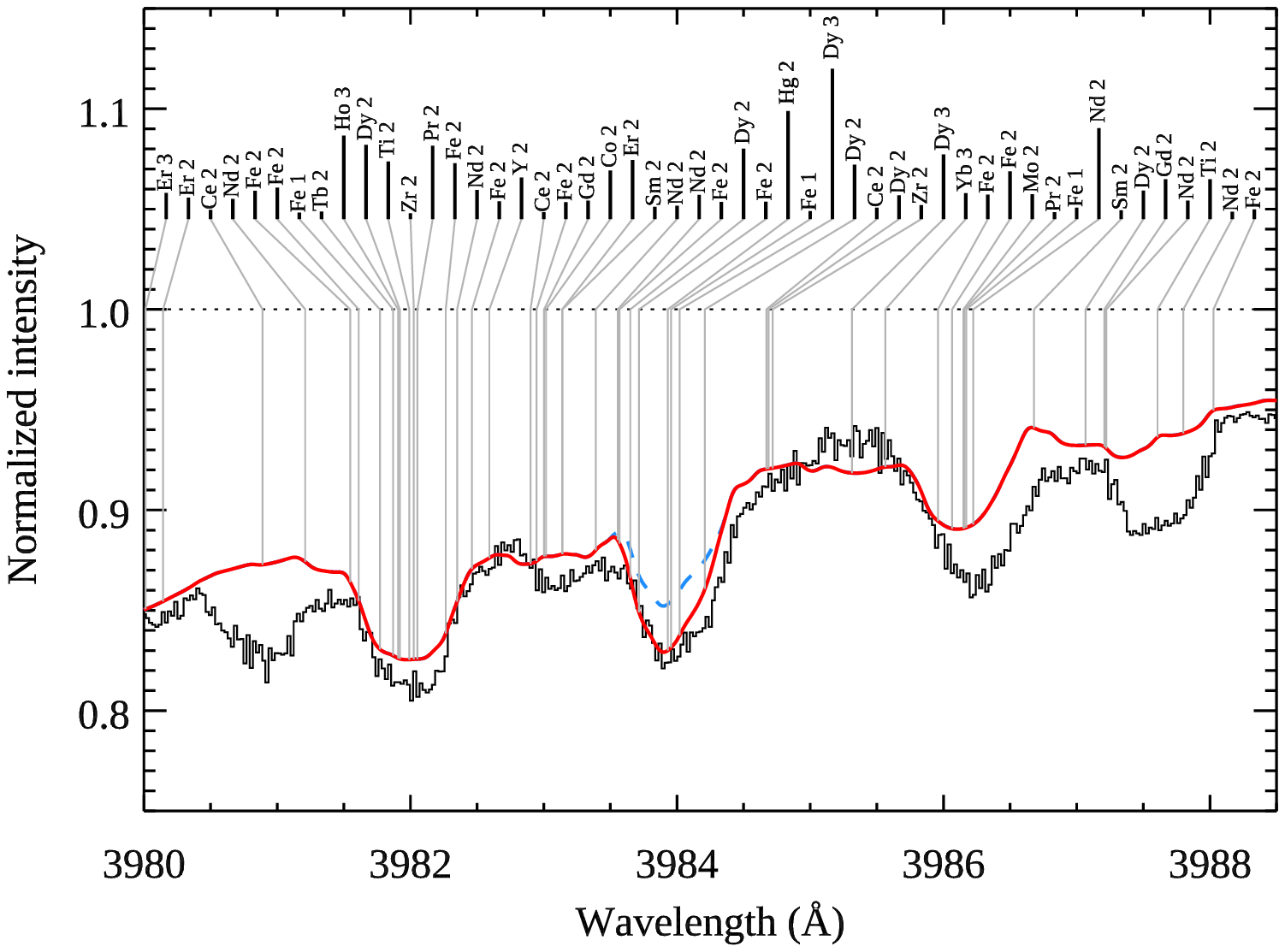}
\caption{Disentangled spectrum of the primary component of \hd\ (black histogram) in the vicinity of the Hg~{\sc ii} 3984~\AA\ line compared to the synthetic spectrum calculated with the abundances from \citetalias{niemczura:2017} (red solid line) and with solar abundance of Hg (blue dashed line). The vertical lines indicate positions of the main spectral features contributing to this wavelength region. The length of the bar below the ion identification indicates the relative line strength.}
\label{fig:hg3984}
\end{figure}

We performed a binary spectral disentangling (separation) using the entire collection of our fourteen ESPaDOnS spectra of \hd. These calculations were carried out for several dozen 100--200~\AA\ wide wavelength regions in the interval between 3940 and 7870~\AA\ with the help of the disentangling method described by \citet{folsom:2010,folsom:2013a}. The radial velocities of both components were fixed to the values measured from LSD profiles (Sect.~\ref{rvlsd}). The average spectra of \hd\ A and B were derived iteratively, using an IDL implementation of the truncated Newton optimisation algorithm. Continuum normalisation of short segments of individual observations was refined simultaneously with the derivation of the component spectra, using quadratic or cubic polynomials.

An application of this disentangling procedure is illustrated in Fig.~\ref{fig:sbfit} for the 4475--4505~\AA\ region. This figure shows individual observed spectra, the final fit obtained by the disentangling code, and the resulting mean, separated spectra of the primary and secondary. As a by-product of this analysis, we also calculated the standard deviation spectra characterising residuals of the fit in the rest frames of the primary and secondary. This information will be used below (Sect.~\ref{lpv}) for the search of intrinsic line profile variability.

The resulting high-quality (S/N ratio of about 500) mean spectrum of the primary component was compared to a theoretical synthetic spectrum for the purpose of obtaining an accurate estimate of the projected rotational velocity, \vs, and verifying the element abundances reported by \citetalias{niemczura:2017}. The synthetic spectrum of \hd\ A was computed with the {\sc Synth3} code for the $T_{\rm eff}=13000$~K, $\log g=4.05$ atmospheric model, zero microturbulent velocity and adopting a line list retrieved from the VALD database. The continuum dilution by the secondary was taken into account by combining the synthetic spectrum of the primary with a featureless continuum flux spectrum of the secondary using the {\sc BinMag} IDL code \footnote{\url{http://www.astro.uu.se/~oleg/binmag.html}}. This software was then employed for analysis of individual wavelength regions and for an interactive adjustment of stellar parameters, including abundances and \vs. 

The projected rotational velocity of the primary, \vs\,=\,$30.5\pm1.1$~\kms, was inferred by fitting about 30 unblended spectral lines. The derived value is consistent with $29.6\pm0.4$~\kms\ expected from the oblique rotator relation assuming synchronisation and alignment of the rotational and orbital axes.

We assess whether our spectra are compatible with the abundance pattern
presented by \citetalias{niemczura:2017}. We do not carry out a complete new abundance analysis since, at this stage,
our central aim is to scrutinise the evidence that the abundance pattern of the primary resembles that of an HgMn-type star. As commonly practiced by spectroscopic studies of these stars \citep{woolf:1999a,monier:2015,monier:2018}, identification of the absorption lines and measurement of large overabundances of certain key elements, such as Xe, Ga, Mn, Au, Pt, Hg, is necessary for assigning the HgMn peculiarity type. However, a conclusive identification and reliable abundance measurement of some of these elements may be challenging for \hd\ A given its relatively rapid rotation and the presence of numerous unidentified rare-earth blends. To this end, we have visually compared our disentangled spectrum of \hd\ A with the synthetic spectrum calculated using the abundances reported by \citetalias{niemczura:2017}.

This assessment leads to mixed results. The previous abundance estimates
of common Fe-peak (Fe, Ti, Cr, Mn), light (Mg, Ca, Si) and rare-earth (Eu, Nd, Pr, Dy, Ho) elements exhibiting many (or a few well-known) strong lines appear to be accurate. The absence of the He~{\sc i} 5876~\AA\ line suggests that helium is more underabundant ([He]\,$\le$\,$-2.2$) than determined by \citetalias{niemczura:2017}. At the same time, large overabundances of many other elements estimated based on a small number of weak spectral features are questionable. For some elements (Ga, Au, Hg), a handful of lines predicted to be observable in the spectrum of the primary appear to be minor contributors to blends with known lines of other elements and/or overlap with broad unidentified features in the observed spectrum likely produced by rare-earth elements. For other elements (Cl, P, S, Xe, Lu, Hf, Pt, Pb), our spectrum synthesis based on published abundances produces unobserved spectral details, suggesting that the reported concentrations of these elements are overestimated by at least 0.5--1.5~dex. Moreover, in many cases no satisfactory fit to observations could be achieved even after implementing abundance reductions in the spectrum synthesis, suggesting the influence of unrecognised blends.

\begin{figure}
\centering
\figps{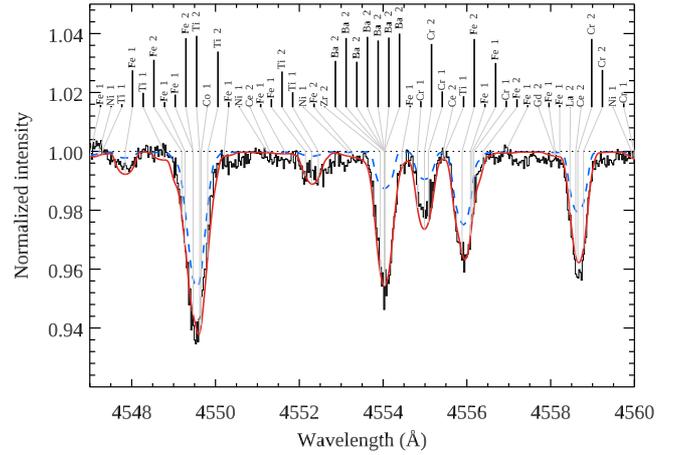}
\caption{Disentangled spectrum of the secondary component of \hd\ (black histogram) compared to the synthetic spectrum calculated with the enhanced Fe-peak and Ba abundances (red solid line) and with solar element abundances (blue dashed line). The vertical lines indicate positions of the main spectral features contributing to this wavelengths region. The length of the bar below the ion identification indicates the relative line strength.}
\label{fig:bspec}
\end{figure}

\begin{figure*}
\centering
\fps{15cm}{90}{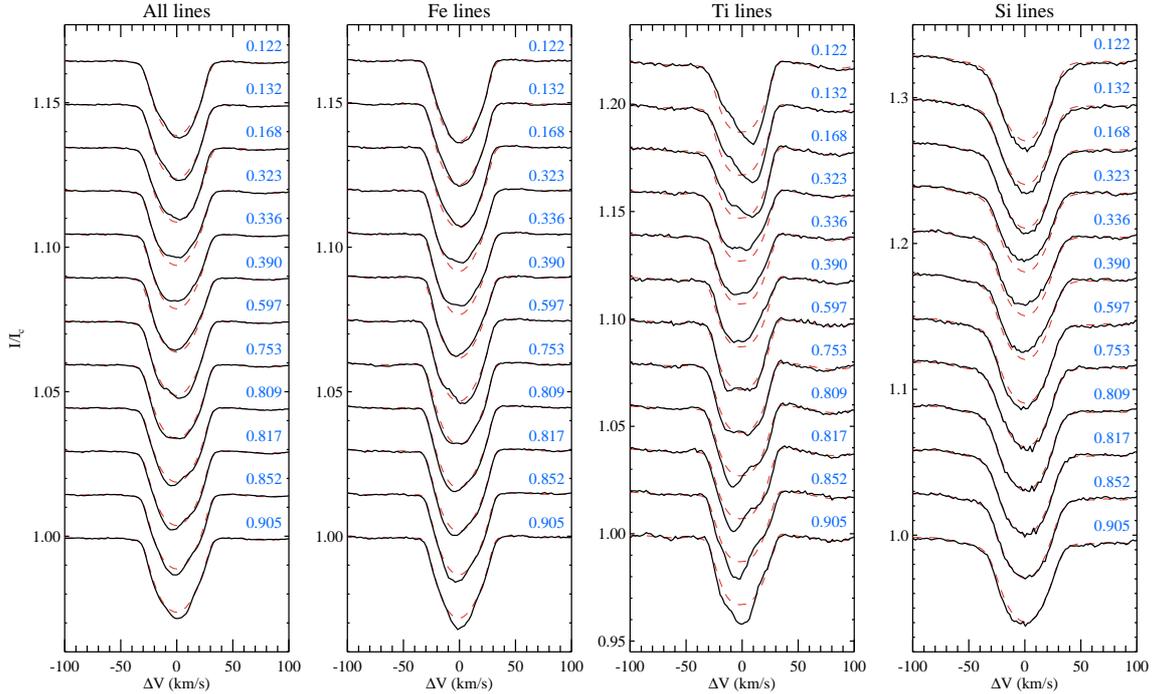}
\caption{Variation of the LSD profiles of \hd\ A derived from the entire set of metal lines as well as from the Fe, Ti, and Si lines. The spectra are offset vertically according to the rotational phase. The time-dependent spectra (black solid curves) are plotted together with the mean profiles (red dashed curves).}
\label{fig:lpv}
\end{figure*}

The ambiguous identification of the Hg~{\sc ii} 3984~\AA\ line, illustrated in Fig.~\ref{fig:hg3984}, is symptomatic of the problems described above. On the one hand, our spectrum synthesis with the reported 3.7~dex mercury overabundance does not contradict the disentangled spectrum of the primary in the vicinity of this Hg~{\sc ii} line. On the other hand, a closer inspection of the synthetic spectrum reveals that the mercury line is a weaker component of the blend with the Dy~{\sc iii} 3984.02~\AA\ feature. In this situation, a determination of the Hg abundance hinges on the knowledge of the doubly ionised Dy spectrum. Moreover, one can also see major discrepancies between the observed and theoretical spectra in the form of broad, unidentified lines at 3981 and 3987~\AA, likely stemming from incomplete and inaccurate rare-earth line lists in this spectral region. In the presence of such artefacts, we do not believe that the reality of the Hg absorption can be ascertained with confidence. No other optical Hg~{\sc i} or {\sc ii} spectral lines are predicted to be deeper than 2--3\% of the continuum (before Doppler and instrumental broadening) by our calculations. 

We carried out similar comparison of the disentangled observations with the synthetic spectra of the secondary, corrected for the continuum dilution by the primary. In this analysis we made use of the $T_{\rm eff}=9000$~K, $\log g =4.40$ model atmosphere and adopted a microturbulent velocity of 2~\kms. It turns out that our initial assumptions of the solar abundance table provides a poor description of the spectrum of the secondary. Based on the analysis of about one hundred spectral features in the 4000--6200~\AA\ wavelength interval we found that observations are best matched with a moderate enhancement or near-solar abundances of light elements ([O]\,=\,+0.25, [Ca]\,=\,0.0, [Si]\,=\,+0.3, [Mg]\,=\,+0.3), a prominent underabundance of carbon and scandium ([C]\,$\le$\,$-0.5$, [Sc]\,=\,$-0.6$), significant overabundance of the Fe-peak elements ([Ti]\,=\,+1.0, [Cr]\,=\,+1.2, [Fe]\,=\,+0.9), and a large enhancement of heavy elements ([Y]\,=\,+1.6, [Sr]\,=\,1.7, [Ba]\,=\,+2.7). Figure~\ref{fig:bspec} illustrates the comparison of the calculations and disentangled spectrum of the secondary for the region around the Ba~{\sc ii} 4554~\AA\ transition. Large overabundances of Fe, Cr, and Ba are readily apparent.

The abundance characteristics of the secondary uncovered by our analysis, in particular a deficiency of Sc combined with an overabundance of Fe-peak and heavy elements, suggests that \hd\ B is a fairly extreme metallic line (Am) star.

The projected rotational velocity of the secondary was found to be $19.1\pm0.9$~\kms, which is higher than \vs\,=\,$14.8\pm0.5$~\kms\ expected for the synchronised rotation.

\subsection{Line profile variability}
\label{lpv}

As was mentioned in Sect.~\ref{abn}, we have investigated intrinsic profile variability of individual spectral lines as part of the binary disentailing procedure. To this end, a correlation of an enhancement of the residuals with line positions in the spectrum of one or both of the binary components points to intrinsic (rotational) spectral variability of that star. As an example of this situation, Fig.~\ref{fig:sbfit} shows peaks in the standard deviation spectrum coinciding with the Mg~{\sc ii} 4481~\AA\ triplet, Ti~{\sc ii} 4488.3, and Ho~{\sc iii} 4494.5~\AA\ lines, and a blend at $\lambda\approx4501$~\AA\ comprising one Nd~{\sc iii} and several Ti~{\sc ii} lines contributed by the primary component of \hd. Using similar analysis across the entire available wavelength range we were able to establish persistent variability of all Ti~{\sc ii} lines and weaker changes in the absorption features of O~{\sc i}, Si~{\sc ii}, Mg~{\sc ii}, Nd~{\sc ii}, Pr~{\sc ii}, Eu~{\sc ii}, and Ho~{\sc iii} for the primary star. No evidence of intrinsic variability was found for any lines of the secondary component.

To obtain further information on these line profile changes, we compared the LSD Stokes $I$ spectra constructed using the full line mask, as discussed in Sect.~\ref{calclsd}, with the LSD profiles obtained for the masks containing only Fe (1056 lines), Ti (99 lines), and Si (74 lines) absorption features. For the latter three calculations we made use of the multi-profile capability of the LSD code described by \citet{kochukhov:2010a} to recover clean mean profiles of a given element, with contributions by all remaining lines taken into account via an independently-derived second LSD profile. The resulting LSD profiles are shown in Fig.~\ref{fig:lpv} for all observations except phases 0.545 and 0.966 for which the line of the primary star is partially blended by the contribution of the secondary. As evident from this figure, a very similar weak variability pattern is present in the LSD profiles obtained with all metal lines and with Fe lines alone. This is not surprising considering that Fe lines dominate the metal line mask. In comparison, the Si LSD profiles exhibit changes of similar magnitude but with a different phase dependence. The strongest spectral changes are seen for the Ti LSD profiles. In this case, the variability pattern is somewhat reminiscent of that of Fe but is more distinct.

\begin{figure}
\centering
\fifps{8cm}{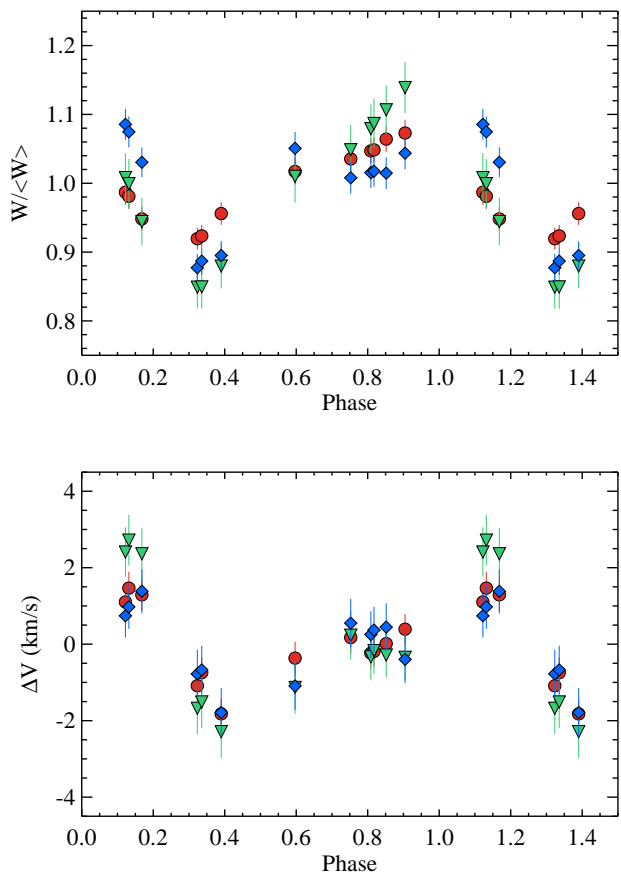}
\caption{Variation of the relative equivalent widths (upper panel) and radial velocities (lower panel) of the Stokes $I$ LSD profiles as a function of rotational phase. The symbols show measurements for Fe (red circles), Ti (green triangles), and Si (blue diamonds).}
\label{fig:ew_rv}
\end{figure}

To quantify the spectral variability of \hd\ A, we studied the equivalent width and centre-of-gravity radial velocities extracted from the set of LSD profiles displayed in Fig.~\ref{fig:lpv}. These measurements are illustrated in Fig.~\ref{fig:ew_rv}. For the strongest variability case (Ti lines), the radial velocity varies by 5.0~\kms\ (peak-to-peak) and the equivalent width changes by 30\%. For Fe and Si, the variability amplitudes are $\approx$\,3~\kms\ and 15--20\% for the radial velocity and equivalent width, respectively. For all three elements considered here both the centre-of-gravity radial velocity and the equivalent width exhibit a coherent variation when phased with $P_{\rm orb}=4.749215$~d, confirming that the axial rotation of \hd\ A is indeed synchronised with its binary orbital motion.

\section{Conclusions and discussion}
\label{disc}

In this study we carried out a detailed analysis of the unique early-type eclipsing binary system \hd. This object was identified by previous photometric studies as an unusual example of a star showing both detached binary eclipses and a smooth out-of-eclipse photometric modulation typical of $\alpha^2$~CVn-type variables. Our high-resolution, time-series spectropolarimetric observations of \hd\ demonstrated the presence of a global, predominantly dipolar magnetic field on the surface of the primary component. No evidence of a magnetic field was found for the secondary. The spectral lines of the primary exhibit variability (in addition to the Doppler shifts due to the orbital motion), both in terms of the equivalent width and line shapes, which is different depending on the considered chemical element. The strongest variability is found for the ionised Ti lines.

These magnetic and spectral variability characteristics suggest that \hd\ A is a magnetic chemically peculiar star with a non-uniform surface distribution of chemical abundances. The chemical spots in such stars are known to produce a flux redistribution between different parts of their spectra, resulting in an inhomogeneous surface brightness distribution at any given wavelength \citep[e.g.][]{krticka:2012}. The rotation of the primary is synchronised with the orbital motion, giving rise to a stable photometric variability superimposed onto the eclipses.

The previous detailed spectroscopic study of \hd\ by \citetalias{niemczura:2017} provided abundance estimates for 46 chemical elements, including many exotic light and heavy species typically identified in the spectra of slowly rotating HgMn stars. Their abundance analysis relied on a single observation of the composite binary spectrum aided by an additional lower-quality spectrum taken during the secondary eclipse. Based on their abundance analysis results and comparison with the surface chemistry of the extreme HgMn star HD\,65949 \citep{cowley:2010}, \citetalias{niemczura:2017} concluded that the primary is an HgMn-related object and that the secondary is likely to be a normal star with solar abundances. 

\begin{figure}
\centering
\figps{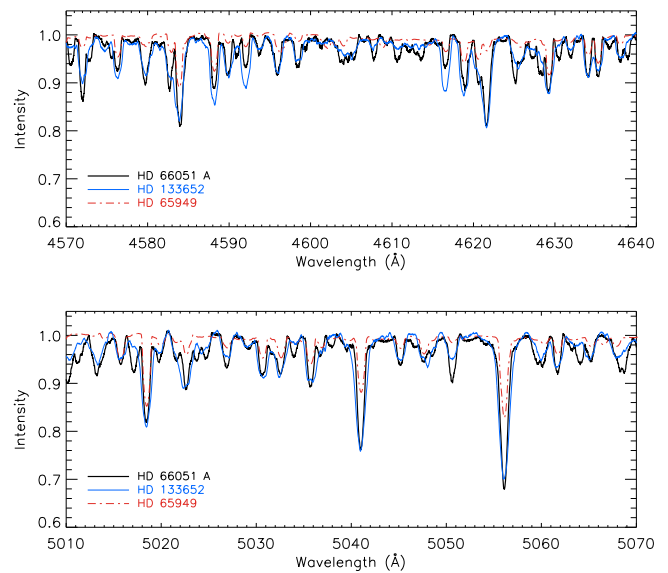}
\caption{Comparison of the disentangled spectrum of \hd\ A (thick solid black line) with the average spectrum of the magnetic Bp star HD\,133652 (thin solid blue line) and the broadened spectrum of the extreme HgMn star HD\,65949 (dashed-dotted red line).}
\label{fig:compare}
\end{figure}

It should be noted that precision and reliability of a stellar abundance analysis quickly diminishes with increasing projected rotational velocity of the star, especially in the presence of a rich and poorly understood rare-earth absorption spectrum. For this reason, typical abundance determinations for rapidly rotating Ap/Bp and HgMn stars \citep[e.g.][]{glagolevskii:2005,semenko:2008,fossati:2011,bailey:2015} are usually restricted to much fewer chemical elements than can be measured for slowly rotating stars of these types \citep[e.g.][]{cowley:2010,shulyak:2010a,castelli:2017}. The large number of chemical elements reported by \citetalias{niemczura:2017}  for \hd\ A deviates significantly from this trend. Their paper provides no details (such as an in-depth discussion of the abundance determination methodology or a list of spectral lines used for abundance measurements) of how the authors have tackled the formidable problem of abundance analysis of this rapidly rotating, rare-earth rich chemically peculiar star.

In the context of the present study, we have acquired 14 individual epochs of spectroscopic observations of \hd. These data were processed with a binary disentangling algorithm, yielding high-quality, separated mean spectra of the primary and secondary. Our spectrum synthesis analysis of the disentangled spectrum of the primary component indicates that reliable line identification is significantly hindered by the rotational broadening and blending by rare-earth lines. Our data are compatible with some of the abundance estimates (e.g. Si, Ca, Fe-peak and rare-earth elements) made by \citetalias{niemczura:2017}. At the same time, we are not able to confirm and, in some cases, we rule out large overabundances claimed for many other elements, including those (Hg, Au, Pt, Xe, Ga) that were used by \citetalias{niemczura:2017} to classify \hd\ A as an HgMn-related star. 

\citetalias{niemczura:2017} suggested that the abundance pattern of \hd\ A is similar to that of the extreme HgMn star HD\,65949. However, a direct comparison of the broadened spectrum of HD\,65949 from the study of \citet{makaganiuk:2011a} with the disentangled spectrum of \hd\ A shows little resemblance (Fig.~\ref{fig:compare}). Despite similar temperatures, the latter star exhibits many more and stronger metal absorption features. On the other hand, He lines are weaker in \hd\ A, indicating an extreme helium deficiency atypical of HgMn stars. In addition, the 1~dex silicon and 4--5~dex rare-earth overabundances are also uncharacteristic of HgMn stars \citep{ghazaryan:2016}. To this end, the spectrum of the primary component of \hd\ appears to be that of a typical silicon and rare-earth rich, magnetic, late-B (Bp) star, in line with its classification by \citet{bidelman:1973} and \citet{houk:1988}. As illustrated by Fig.~\ref{fig:compare}, this conclusion is supported by the similarity of the line absorption in our disentangled spectrum of \hd\ A with the spectrum of unremarkable magnetic Bp star HD\,133652 \citep{bailey:2015}.

Our definite detection of a global magnetic field in \hd\ A also clearly sets this object aside from HgMn stars. The latter are thought to be non-magnetic \citep{shorlin:2002,auriere:2010a,makaganiuk:2011a,makaganiuk:2012,kochukhov:2011b,kochukhov:2013a} at the 10--100~G level. Occasional claims of statistically significant longitudinal field detections for HgMn stars can be found in the literature \citep{hubrig:2006b,hubrig:2010,hubrig:2012}, but those have not been confirmed by independent follow-up studies, including re-analyses of the same observational data \citep{bagnulo:2012,kochukhov:2013a}, and were never supported by direct Zeeman polarisation signature detections such as that presented here for \hd\ A.

The high S/N ratio disentangled spectrum of the faint secondary enabled a new insight into its nature. The projected rotational velocity of the secondary suggests that its rotational period is shorter that the orbital period of the system or that its rotational axis is misaligned with the orbital axis. The spectrum of the secondary shows enhanced lines of Fe-peak elements, a deficiency of Sc and large overabundances of Sr, Y, and Ba. These abundance characteristics, as well as the lack of global magnetic field and line profile variability, are typical of metallic-line (Am) stars. This establishes \hd\ as belonging to the group of several confirmed systems hosting both an Ap/Bp and Am components. The three other binaries in this group are HD\,98088, HD\,5550, and BD-19\,5044L \citep{folsom:2013a,alecian:2016,landstreet:2017}. Together with \hd, these systems represent all well-studied close SB2s with lower-mass ($M \le 4 M_\odot$) magnetic Ap/Bp primaries.

We used archival broad-band photometric observations of \hd\ together with the radial velocities derived from our data to perform binary system modelling with the {\sc PHOEBE} code. This analysis provided precise fundamental radii and masses of both components. In addition, information on the wavelength-dependent luminosity ratio combined with the results of hydrogen line profile modelling allowed us to infer effective temperatures of both stars. 

Our binary modelling results can be compared with the parameters of the system recently published by \citet{paunzen:2018}. These authors used an earlier version of the {\sc PHOEBE} code for the analysis of the same photometric light curves as investigated here. On the other hand, they relied on an independent set of (unpublished) radial velocity estimates derived from heterogeneous spectroscopic observations, including measurements from medium-resolution ($R$\,$\sim$\,12000) spectra. This RV dataset has phase gaps of up to about 1/3 of the orbital cycle. In comparison, the largest phase gap in our spectral time series is 0.156. Furthermore, \citet{paunzen:2018} treated their RV measurements with a separate analysis, from which they derived the mass ratio and then fixed it in the {\sc PHOEBE} solution. Their study also lacked the rigorous Bayesian MCMC error assessment comparable to that presented in our paper. Likely owing to these differences, the stellar radii ($R_1=2.58\pm0.13 R_\odot$ and $R_2=1.61\pm0.10 R_\odot$) obtained by \citet{paunzen:2018} are marginally (at a 1.5--2.0$\sigma$ level) different and less precise compared to our results. Their inclination angle ($i=84.7\pm0.1\degr$) is significantly lower than our estimate. On the other hand, their mass ratio ($M_2/M_1 = 0.56\pm0.03$) and individual component masses ($M_1=3.23\pm0.22 M_\odot$ and $M_2=1.81\pm0.13 R_\odot$) agree with our results within error bars.

The set of observational constraints available for \hd\ is unprecedented for early-type stars with such extreme surface chemical abundance anomalies and magnetic field. These results open unique possibilities for detailed stellar interior structure and evolution studies. For example, one could exploit \hd\ to test stellar structure models with radiatively-driven chemical stratification \citep{vick:2010} and put firm limits on the radius modification due to interior non-force-free magnetic fields \citep{valyavin:2004}. One could also probe other poorly understood interior structure processes such as mixing in the radiative zone, convective core overshoot, etc.

Our study of the \hd\ system also offers a valuable new insight into the long-standing problem of the interplay between fossil magnetism and binarity.  This object belongs to an exclusive group of about ten well-characterised close ($P_{\rm orb} < 20$~d) SB2 systems with an early-type magnetic component \citep{landstreet:2017,shultz:2018}. Two less well-studied close SB1 binaries, HD\,25267 and HD 25823, with an early-type magnetic component are also known \citep{mathys:2017}. Seven of these SB1 and SB2 systems, including \hd, have $P_{\rm orb} < 10$~d. Their existence defies the well-established general trend of a lack of intermediate-mass and massive magnetic stars in close binaries and challenges theories purporting to explain this trend \citep{commercon:2011,schneider:2016}. Evidently, these theories need to account for the formation of at least a few close systems with early-type magnetic components. Among these key objects, \hd\ is the only known eclipsing binary and hence the only system with fundamental constraints on the component parameters and their dynamical interaction.

\section*{Acknowledgments} 
OK acknowledges financial support from the Knut and Alice Wallenberg Foundation, the Swedish Research Council, and the Swedish National Space Board. 
Part of the research leading to these results has received funding from the European Research Council (ERC) under the European Union's Horizon 2020 research and innovation programme (grant agreement N$^\circ$670519: MAMSIE). The computational resources and services used in this work were in part provided by the VSC (Flemish Supercomputer Center), funded by the Research Foundation -- Flanders (FWO) and the Flemish Government -- department EWI.
GAW acknowledges Discovery Grant support from the Natural Sciences and Engineering Research Council (NSERC) of Canada.


\begin{thebibliography}{}
\makeatletter
\relax
\def\mn@urlcharsother{\let\do\@makeother \do\$\do\&\do\#\do\^\do\_\do\%\do\~}
\def\mn@doi{\begingroup\mn@urlcharsother \@ifnextchar [ {\mn@doi@}
  {\mn@doi@[]}}
\def\mn@doi@[#1]#2{\def\@tempa{#1}\ifx\@tempa\@empty \href
  {http://dx.doi.org/#2} {doi:#2}\else \href {http://dx.doi.org/#2} {#1}\fi
  \endgroup}
\def\mn@eprint#1#2{\mn@eprint@#1:#2::\@nil}
\def\mn@eprint@arXiv#1{\href {http://arxiv.org/abs/#1} {{\tt arXiv:#1}}}
\def\mn@eprint@dblp#1{\href {http://dblp.uni-trier.de/rec/bibtex/#1.xml}
  {dblp:#1}}
\def\mn@eprint@#1:#2:#3:#4\@nil{\def\@tempa {#1}\def\@tempb {#2}\def\@tempc
  {#3}\ifx \@tempc \@empty \let \@tempc \@tempb \let \@tempb \@tempa \fi \ifx
  \@tempb \@empty \def\@tempb {arXiv}\fi \@ifundefined
  {mn@eprint@\@tempb}{\@tempb:\@tempc}{\expandafter \expandafter \csname
  mn@eprint@\@tempb\endcsname \expandafter{\@tempc}}}

\bibitem[\protect\citeauthoryear{{Achilleos} \& {Wickramasinghe}}{{Achilleos}
  \& {Wickramasinghe}}{1989}]{achilleos:1989}
{Achilleos} N.,  {Wickramasinghe} D.~T.,  1989, \mn@doi [\apj]
  {10.1086/168024}, \href {http://adsabs.harvard.edu/abs/1989ApJ...346..444A}
  {346, 444}

\bibitem[\protect\citeauthoryear{{Adelman}, {Gulliver}, {Kochukhov}  \&
  {Ryabchikova}}{{Adelman} et~al.}{2002}]{adelman:2002}
{Adelman} S.~J.,  {Gulliver} A.~F.,  {Kochukhov} O.~P.,   {Ryabchikova} T.~A.,
  2002, \mn@doi [\apj] {10.1086/341140}, \href
  {http://adsabs.harvard.edu/abs/2002ApJ...575..449A} {575, 449}

\bibitem[\protect\citeauthoryear{{Alecian} et~al.,}{{Alecian}
  et~al.}{2008}]{alecian:2008a}
{Alecian} E.,  et~al., 2008, \mn@doi [\mnras]
  {10.1111/j.1365-2966.2008.12842.x}, \href
  {http://adsabs.harvard.edu/abs/2008MNRAS.385..391A} {385, 391}

\bibitem[\protect\citeauthoryear{{Alecian} et~al.,}{{Alecian}
  et~al.}{2015}]{alecian:2015a}
{Alecian} E.,  et~al., 2015, in {Meynet} G.,  {Georgy} C.,  {Groh} J.,   {Stee}
  P.,  eds,  IAU Symposium Vol. 307, New Windows on Massive Stars. pp 330--335

\bibitem[\protect\citeauthoryear{{Alecian}, {Tkachenko}, {Neiner}, {Folsom}  \&
  {Leroy}}{{Alecian} et~al.}{2016}]{alecian:2016}
{Alecian} E.,  {Tkachenko} A.,  {Neiner} C.,  {Folsom} C.~P.,   {Leroy} B.,
  2016, \mn@doi [\aap] {10.1051/0004-6361/201527355}, \href
  {http://adsabs.harvard.edu/abs/2016A%26A...589A..47A} {589, A47}

\bibitem[\protect\citeauthoryear{{Asplund}, {Grevesse}, {Sauval}  \&
  {Scott}}{{Asplund} et~al.}{2009}]{asplund:2009}
{Asplund} M.,  {Grevesse} N.,  {Sauval} A.~J.,   {Scott} P.,  2009, \mn@doi
  [\araa] {10.1146/annurev.astro.46.060407.145222}, \href
  {http://adsabs.harvard.edu/abs/2009ARA%26A..47..481A} {47, 481}

\bibitem[\protect\citeauthoryear{{Auri{\`e}re} et~al.,}{{Auri{\`e}re}
  et~al.}{2010}]{auriere:2010a}
{Auri{\`e}re} M.,  et~al., 2010, \mn@doi [\aap] {10.1051/0004-6361/201014848},
  \href {http://adsabs.harvard.edu/abs/2010A%26A...523A..40A} {523, A40}

\bibitem[\protect\citeauthoryear{{Bagnulo}, {Landi degl'Innocenti}  \& {Landi
  degl'Innocenti}}{{Bagnulo} et~al.}{1996}]{bagnulo:1996}
{Bagnulo} S.,  {Landi degl'Innocenti} M.,   {Landi degl'Innocenti} E.,  1996,
  \aap, \href {http://adsabs.harvard.edu/abs/1996A%26A...308..115B} {308, 115}

\bibitem[\protect\citeauthoryear{{Bagnulo}, {Landolfi}, {Landstreet}, {Landi
  Degl'Innocenti}, {Fossati}  \& {Sterzik}}{{Bagnulo}
  et~al.}{2009}]{bagnulo:2009}
{Bagnulo} S.,  {Landolfi} M.,  {Landstreet} J.~D.,  {Landi Degl'Innocenti} E.,
  {Fossati} L.,   {Sterzik} M.,  2009, \mn@doi [\pasp] {10.1086/605654}, \href
  {http://adsabs.harvard.edu/abs/2009PASP..121..993B} {121, 993}

\bibitem[\protect\citeauthoryear{{Bagnulo}, {Landstreet}, {Fossati}  \&
  {Kochukhov}}{{Bagnulo} et~al.}{2012}]{bagnulo:2012}
{Bagnulo} S.,  {Landstreet} J.~D.,  {Fossati} L.,   {Kochukhov} O.,  2012,
  \mn@doi [\aap] {10.1051/0004-6361/201118098}, \href
  {http://adsabs.harvard.edu/abs/2012A%26A...538A.129B} {538, A129}

\bibitem[\protect\citeauthoryear{{Bailey} \& {Landstreet}}{{Bailey} \&
  {Landstreet}}{2015}]{bailey:2015}
{Bailey} J.~D.,  {Landstreet} J.~D.,  2015, \mn@doi [\aap]
  {10.1051/0004-6361/201526073}, \href
  {http://adsabs.harvard.edu/abs/2015A%26A...580A..81B} {580, A81}

\bibitem[\protect\citeauthoryear{{Bidelman} \& {MacConnell}}{{Bidelman} \&
  {MacConnell}}{1973}]{bidelman:1973}
{Bidelman} W.~P.,  {MacConnell} D.~J.,  1973, \aj, \href
  {http://cdsads.u-strasbg.fr/abs/1973AJ.....78..687B} {78, 687}

\bibitem[\protect\citeauthoryear{{Braithwaite} \& {Spruit}}{{Braithwaite} \&
  {Spruit}}{2004}]{braithwaite:2004}
{Braithwaite} J.,  {Spruit} H.~C.,  2004, \mn@doi [\nat] {10.1038/nature02934},
  \href {http://adsabs.harvard.edu/abs/2004Natur.431..819B} {431, 819}

\bibitem[\protect\citeauthoryear{{Carrier}, {North}, {Udry}  \&
  {Babel}}{{Carrier} et~al.}{2002}]{carrier:2002}
{Carrier} F.,  {North} P.,  {Udry} S.,   {Babel} J.,  2002, \mn@doi [\aap]
  {10.1051/0004-6361:20021122}, \href
  {http://adsabs.harvard.edu/abs/2002A%26A...394..151C} {394, 151}

\bibitem[\protect\citeauthoryear{{Castelli}, {Cowley}, {Ayres}, {Catanzaro}  \&
  {Leone}}{{Castelli} et~al.}{2017}]{castelli:2017}
{Castelli} F.,  {Cowley} C.~R.,  {Ayres} T.~R.,  {Catanzaro} G.,   {Leone} F.,
  2017, \mn@doi [\aap] {10.1051/0004-6361/201629854}, \href
  {http://adsabs.harvard.edu/abs/2017A%26A...601A.119C} {601, A119}

\bibitem[\protect\citeauthoryear{{Claret} \& {Bloemen}}{{Claret} \&
  {Bloemen}}{2011}]{claret:2011}
{Claret} A.,  {Bloemen} S.,  2011, \mn@doi [\aap]
  {10.1051/0004-6361/201116451}, \href
  {http://adsabs.harvard.edu/abs/2011A%26A...529A..75C} {529, A75}

\bibitem[\protect\citeauthoryear{{Commer{\c c}on}, {Hennebelle}  \&
  {Henning}}{{Commer{\c c}on} et~al.}{2011}]{commercon:2011}
{Commer{\c c}on} B.,  {Hennebelle} P.,   {Henning} T.,  2011, \mn@doi [\apjl]
  {10.1088/2041-8205/742/1/L9}, \href
  {http://adsabs.harvard.edu/abs/2011ApJ...742L...9C} {742, L9}

\bibitem[\protect\citeauthoryear{{Cowley}, {Hubrig}, {Palmeri}, {Quinet},
  {Bi{\'e}mont}, {Wahlgren}, {Sch{\"u}tz}  \& {Gonz{\'a}lez}}{{Cowley}
  et~al.}{2010}]{cowley:2010}
{Cowley} C.~R.,  {Hubrig} S.,  {Palmeri} P.,  {Quinet} P.,  {Bi{\'e}mont}
  {\'E}.,  {Wahlgren} G.~M.,  {Sch{\"u}tz} O.,   {Gonz{\'a}lez} J.~F.,  2010,
  \mn@doi [\mnras] {10.1111/j.1365-2966.2010.16529.x}, \href
  {http://cdsads.u-strasbg.fr/abs/2010MNRAS.405.1271C} {405, 1271}

\bibitem[\protect\citeauthoryear{{Donati}, {Semel}  \& {Rees}}{{Donati}
  et~al.}{1992}]{donati:1992}
{Donati} J.-F.,  {Semel} M.,   {Rees} D.~E.,  1992, \aap, \href
  {http://adsabs.harvard.edu/abs/1992A%26A...265..669D} {265, 669}

\bibitem[\protect\citeauthoryear{{Donati}, {Semel}, {Carter}, {Rees}  \&
  {Collier Cameron}}{{Donati} et~al.}{1997}]{donati:1997}
{Donati} J.-F.,  {Semel} M.,  {Carter} B.~D.,  {Rees} D.~E.,   {Collier
  Cameron} A.,  1997, \mnras, \href
  {http://adsabs.harvard.edu/abs/1997MNRAS.291..658D} {291, 658}

\bibitem[\protect\citeauthoryear{{Folsom}, {Kochukhov}, {Wade}, {Silvester}  \&
  {Bagnulo}}{{Folsom} et~al.}{2010}]{folsom:2010}
{Folsom} C.~P.,  {Kochukhov} O.,  {Wade} G.~A.,  {Silvester} J.,   {Bagnulo}
  S.,  2010, \mn@doi [\mnras] {10.1111/j.1365-2966.2010.17057.x}, \href
  {http://adsabs.harvard.edu/abs/2010MNRAS.407.2383F} {407, 2383}

\bibitem[\protect\citeauthoryear{{Folsom}, {Wade}  \& {Alecian}}{{Folsom}
  et~al.}{2013a}]{folsom:2013b}
{Folsom} C.~P.,  {Wade} G.~A.,   {Alecian} E.,  2013a, in {Pavlovski} K.,
  {Tkachenko} A.,   {Torres} G.,  eds,  EAS Publications Series Vol. 64, EAS
  Publications Series. pp 119--126

\bibitem[\protect\citeauthoryear{{Folsom}, {Likuski}, {Wade}, {Kochukhov},
  {Alecian}  \& {Shulyak}}{{Folsom} et~al.}{2013b}]{folsom:2013a}
{Folsom} C.~P.,  {Likuski} K.,  {Wade} G.~A.,  {Kochukhov} O.,  {Alecian} E.,
  {Shulyak} D.,  2013b, \mn@doi [\mnras] {10.1093/mnras/stt269}, \href
  {http://adsabs.harvard.edu/abs/2013MNRAS.431.1513F} {431, 1513}

\bibitem[\protect\citeauthoryear{{Foreman-Mackey}, {Hogg}, {Lang}  \&
  {Goodman}}{{Foreman-Mackey} et~al.}{2013}]{foreman-mackey:2013}
{Foreman-Mackey} D.,  {Hogg} D.~W.,  {Lang} D.,   {Goodman} J.,  2013, \mn@doi
  [\pasp] {10.1086/670067}, \href
  {http://adsabs.harvard.edu/abs/2013PASP..125..306F} {125, 306}

\bibitem[\protect\citeauthoryear{{Fossati}, {Folsom}, {Bagnulo}, {Grunhut},
  {Kochukhov}, {Landstreet}, {Paladini}  \& {Wade}}{{Fossati}
  et~al.}{2011}]{fossati:2011}
{Fossati} L.,  {Folsom} C.~P.,  {Bagnulo} S.,  {Grunhut} J.~H.,  {Kochukhov}
  O.,  {Landstreet} J.~D.,  {Paladini} C.,   {Wade} G.~A.,  2011, \mn@doi
  [\mnras] {10.1111/j.1365-2966.2011.18199.x}, \href
  {http://adsabs.harvard.edu/abs/2011MNRAS.413.1132F} {413, 1132}

\bibitem[\protect\citeauthoryear{{Gerbaldi}, {Floquet}  \& {Hauck}}{{Gerbaldi}
  et~al.}{1985}]{gerbaldi:1985}
{Gerbaldi} M.,  {Floquet} M.,   {Hauck} B.,  1985, \aap, \href
  {http://adsabs.harvard.edu/abs/1985A%26A...146..341G} {146, 341}

\bibitem[\protect\citeauthoryear{{Ghazaryan} \& {Alecian}}{{Ghazaryan} \&
  {Alecian}}{2016}]{ghazaryan:2016}
{Ghazaryan} S.,  {Alecian} G.,  2016, \mn@doi [\mnras] {10.1093/mnras/stw911},
  \href {http://adsabs.harvard.edu/abs/2016MNRAS.460.1912G} {460, 1912}

\bibitem[\protect\citeauthoryear{{Glagolevskii}, {Ryabchikova}  \&
  {Chountonov}}{{Glagolevskii} et~al.}{2005}]{glagolevskii:2005}
{Glagolevskii} Y.~V.,  {Ryabchikova} T.~A.,   {Chountonov} G.~A.,  2005,
  \mn@doi [Astronomy Letters] {10.1134/1.1922531}, \href
  {http://adsabs.harvard.edu/abs/2005AstL...31..327G} {31, 327}

\bibitem[\protect\citeauthoryear{{Houk} \& {Smith-Moore}}{{Houk} \&
  {Smith-Moore}}{1988}]{houk:1988}
{Houk} N.,  {Smith-Moore} M.,  1988, {Michigan Catalogue of Two-dimensional
  Spectral Types for the HD Stars. Volume 4, Declinations -26 deg to -12 deg}

\bibitem[\protect\citeauthoryear{{Hubrig}, {North}, {Sch{\"o}ller}  \&
  {Mathys}}{{Hubrig} et~al.}{2006}]{hubrig:2006b}
{Hubrig} S.,  {North} P.,  {Sch{\"o}ller} M.,   {Mathys} G.,  2006, \mn@doi
  [Astronomische Nachrichten] {10.1002/asna.200610535}, \href
  {http://adsabs.harvard.edu/abs/2006AN....327..289H} {327, 289}

\bibitem[\protect\citeauthoryear{{Hubrig} et~al.,}{{Hubrig}
  et~al.}{2010}]{hubrig:2010}
{Hubrig} S.,  et~al., 2010, \mn@doi [\mnras]
  {10.1111/j.1745-3933.2010.00928.x}, \href
  {http://adsabs.harvard.edu/abs/2010MNRAS.408L..61H} {408, L61}

\bibitem[\protect\citeauthoryear{{Hubrig} et~al.,}{{Hubrig}
  et~al.}{2012}]{hubrig:2012}
{Hubrig} S.,  et~al., 2012, \mn@doi [\aap] {10.1051/0004-6361/201219778}, \href
  {http://adsabs.harvard.edu/abs/2012A%26A...547A..90H} {547, A90}

\bibitem[\protect\citeauthoryear{{H{\"u}mmerich}, {Paunzen}  \&
  {Bernhard}}{{H{\"u}mmerich} et~al.}{2016}]{hummerich:2016}
{H{\"u}mmerich} S.,  {Paunzen} E.,   {Bernhard} K.,  2016, \mn@doi [\aj]
  {10.3847/0004-6256/152/4/104}, \href
  {http://adsabs.harvard.edu/abs/2016AJ....152..104H} {152, 104}

\bibitem[\protect\citeauthoryear{{Kochukhov}}{{Kochukhov}}{2007}]{kochukhov:2007d}
{Kochukhov} O.,  2007, in {Romanyuk} I.~I.,  {Kudryavtsev} D.~O.,  eds, Physics
  of Magnetic Stars. pp 109--118

\bibitem[\protect\citeauthoryear{{Kochukhov}, {Piskunov}, {Sachkov}  \&
  {Kudryavtsev}}{{Kochukhov} et~al.}{2005}]{kochukhov:2005b}
{Kochukhov} O.,  {Piskunov} N.,  {Sachkov} M.,   {Kudryavtsev} D.,  2005,
  \mn@doi [\aap] {10.1051/0004-6361:20053123}, \href
  {http://adsabs.harvard.edu/abs/2005A%26A...439.1093K} {439, 1093}

\bibitem[\protect\citeauthoryear{{Kochukhov}, {Makaganiuk}  \&
  {Piskunov}}{{Kochukhov} et~al.}{2010}]{kochukhov:2010a}
{Kochukhov} O.,  {Makaganiuk} V.,   {Piskunov} N.,  2010, \mn@doi [\aap]
  {10.1051/0004-6361/201015429}, \href
  {http://adsabs.harvard.edu/abs/2010A%26A...524A...5K} {524, A5}

\bibitem[\protect\citeauthoryear{{Kochukhov} et~al.,}{{Kochukhov}
  et~al.}{2011}]{kochukhov:2011b}
{Kochukhov} O.,  et~al., 2011, \mn@doi [\aap] {10.1051/0004-6361/201117970},
  \href {http://cdsads.u-strasbg.fr/abs/2011A%26A...534L..13K} {534, L13}

\bibitem[\protect\citeauthoryear{{Kochukhov} et~al.,}{{Kochukhov}
  et~al.}{2013}]{kochukhov:2013a}
{Kochukhov} O.,  et~al., 2013, \mn@doi [\aap] {10.1051/0004-6361/201321467},
  \href {http://adsabs.harvard.edu/abs/2013A%26A...554A..61K} {554, A61}

\bibitem[\protect\citeauthoryear{{Kochukhov}, {L{\"u}ftinger}, {Neiner},
  {Alecian}  \& {MiMeS Collaboration}}{{Kochukhov}
  et~al.}{2014}]{kochukhov:2014}
{Kochukhov} O.,  {L{\"u}ftinger} T.,  {Neiner} C.,  {Alecian} E.,   {MiMeS
  Collaboration} 2014, \mn@doi [\aap] {10.1051/0004-6361/201423472}, \href
  {http://adsabs.harvard.edu/abs/2014A%26A...565A..83K} {565, A83}

\bibitem[\protect\citeauthoryear{{Kochukhov}, {Silvester}, {Bailey},
  {Landstreet}  \& {Wade}}{{Kochukhov} et~al.}{2017}]{kochukhov:2017a}
{Kochukhov} O.,  {Silvester} J.,  {Bailey} J.~D.,  {Landstreet} J.~D.,   {Wade}
  G.~A.,  2017, \mn@doi [\aap] {10.1051/0004-6361/201730919}, \href
  {http://adsabs.harvard.edu/abs/2017A%26A...605A..13K} {605, A13}

\bibitem[\protect\citeauthoryear{{Korhonen} et~al.,}{{Korhonen}
  et~al.}{2013}]{korhonen:2013}
{Korhonen} H.,  et~al., 2013, \mn@doi [\aap] {10.1051/0004-6361/201220951},
  \href {http://adsabs.harvard.edu/abs/2013A%26A...553A..27K} {553, A27}

\bibitem[\protect\citeauthoryear{{Krti{\v c}ka}, {Mikul{\'a}{\v s}ek},
  {L{\"u}ftinger}, {Shulyak}, {Zverko}, {{\v Z}i{\v z}{\v n}ovsk{\'y}}  \&
  {Sokolov}}{{Krti{\v c}ka} et~al.}{2012}]{krticka:2012}
{Krti{\v c}ka} J.,  {Mikul{\'a}{\v s}ek} Z.,  {L{\"u}ftinger} T.,  {Shulyak}
  D.,  {Zverko} J.,  {{\v Z}i{\v z}{\v n}ovsk{\'y}} J.,   {Sokolov} N.~A.,
  2012, \mn@doi [\aap] {10.1051/0004-6361/201117490}, \href
  {http://adsabs.harvard.edu/abs/2012A%26A...537A..14K} {537, A14}

\bibitem[\protect\citeauthoryear{{Kurtz} \& {Martinez}}{{Kurtz} \&
  {Martinez}}{2000}]{kurtz:2000}
{Kurtz} D.~W.,  {Martinez} P.,  2000, Baltic Astronomy, \href
  {http://adsabs.harvard.edu/abs/2000BaltA...9..253K} {9, 253}

\bibitem[\protect\citeauthoryear{{Landi Degl'Innocenti} \& {Landolfi}}{{Landi
  Degl'Innocenti} \& {Landolfi}}{2004}]{polarization:2004}
{Landi Degl'Innocenti} E.,  {Landolfi} M.,  2004, {Polarization in Spectral
  Lines}.
 Astrophysics and Space Science Library Vol. 307, Kluwer Academic Publishers

\bibitem[\protect\citeauthoryear{{Landstreet}}{{Landstreet}}{1988}]{landstreet:1988}
{Landstreet} J.~D.,  1988, \mn@doi [\apj] {10.1086/166155}, \href
  {http://adsabs.harvard.edu/abs/1988ApJ...326..967L} {326, 967}

\bibitem[\protect\citeauthoryear{{Landstreet}, {Kochukhov}, {Alecian},
  {Bailey}, {Mathis}, {Neiner}, {Wade}  \& {BINaMIcS
  Collaboration}}{{Landstreet} et~al.}{2017}]{landstreet:2017}
{Landstreet} J.~D.,  {Kochukhov} O.,  {Alecian} E.,  {Bailey} J.~D.,  {Mathis}
  S.,  {Neiner} C.,  {Wade} G.~A.,   {BINaMIcS Collaboration} 2017, \mn@doi
  [\aap] {10.1051/0004-6361/201630233}, \href
  {http://adsabs.harvard.edu/abs/2017A%26A...601A.129L} {601, A129}

\bibitem[\protect\citeauthoryear{{Makaganiuk} et~al.,}{{Makaganiuk}
  et~al.}{2011a}]{makaganiuk:2011a}
{Makaganiuk} V.,  et~al., 2011a, \mn@doi [\aap] {10.1051/0004-6361/201015666},
  \href {http://adsabs.harvard.edu/abs/2011A%26A...525A..97M} {525, A97}

\bibitem[\protect\citeauthoryear{{Makaganiuk} et~al.,}{{Makaganiuk}
  et~al.}{2011b}]{makaganiuk:2011}
{Makaganiuk} V.,  et~al., 2011b, \mn@doi [\aap] {10.1051/0004-6361/201016302},
  \href {http://adsabs.harvard.edu/abs/2011A%26A...529A.160M} {529, A160}

\bibitem[\protect\citeauthoryear{{Makaganiuk} et~al.,}{{Makaganiuk}
  et~al.}{2012}]{makaganiuk:2012}
{Makaganiuk} V.,  et~al., 2012, \mn@doi [\aap] {10.1051/0004-6361/201118167},
  \href {http://adsabs.harvard.edu/abs/2012A%26A...539A.142M} {539, A142}

\bibitem[\protect\citeauthoryear{{Mann} \& {von Braun}}{{Mann} \& {von
  Braun}}{2015}]{mann:2015}
{Mann} A.~W.,  {von Braun} K.,  2015, \mn@doi [\pasp] {10.1086/680012}, \href
  {http://adsabs.harvard.edu/abs/2015PASP..127..102M} {127, 102}

\bibitem[\protect\citeauthoryear{{Markwardt}}{{Markwardt}}{2009}]{markwardt:2009}
{Markwardt} C.~B.,  2009, in {D.~A.~Bohlender, D.~Durand, \& P.~Dowler} ed.,
  Astronomical Society of the Pacific Conference Series Vol. 411, Astronomical
  Society of the Pacific Conference Series. pp 251--254

\bibitem[\protect\citeauthoryear{{Mathys}}{{Mathys}}{2017}]{mathys:2017}
{Mathys} G.,  2017, \mn@doi [\aap] {10.1051/0004-6361/201628429}, \href
  {http://adsabs.harvard.edu/abs/2017A%26A...601A..14M} {601, A14}

\bibitem[\protect\citeauthoryear{{McLaughlin}}{{McLaughlin}}{1924}]{mclaughlin:1924}
{McLaughlin} D.~B.,  1924, \mn@doi [\apj] {10.1086/142826}, \href
  {http://adsabs.harvard.edu/abs/1924ApJ....60...22M} {60, 22}

\bibitem[\protect\citeauthoryear{{Monier}, {Gebran}  \& {Royer}}{{Monier}
  et~al.}{2015}]{monier:2015}
{Monier} R.,  {Gebran} M.,   {Royer} F.,  2015, \mn@doi [\aap]
  {10.1051/0004-6361/201526106}, \href
  {http://adsabs.harvard.edu/abs/2015A%26A...577A..96M} {577, A96}

\bibitem[\protect\citeauthoryear{{Monier}, {Gebran}, {Royer}, {Kilicoglu}  \&
  {Fr{\'e}mat}}{{Monier} et~al.}{2018}]{monier:2018}
{Monier} R.,  {Gebran} M.,  {Royer} F.,  {Kilicoglu} T.,   {Fr{\'e}mat} Y.,
  2018, \mn@doi [\apj] {10.3847/1538-4357/aaa246}, \href
  {http://adsabs.harvard.edu/abs/2018ApJ...854...50M} {854, 50}

\bibitem[\protect\citeauthoryear{{Morel} et~al.,}{{Morel}
  et~al.}{2014}]{morel:2014}
{Morel} T.,  et~al., 2014, \mn@doi [\aap] {10.1051/0004-6361/201322289}, \href
  {http://adsabs.harvard.edu/abs/2014A%26A...561A..35M} {561, A35}

\bibitem[\protect\citeauthoryear{{Neiner}, {Mathis}, {Alecian}, {Emeriau},
  {Grunhut}, {BinaMIcS}  \& {MiMeS Collaborations}}{{Neiner}
  et~al.}{2015}]{neiner:2015}
{Neiner} C.,  {Mathis} S.,  {Alecian} E.,  {Emeriau} C.,  {Grunhut} J.,
  {BinaMIcS}  {MiMeS Collaborations} 2015, in {Nagendra} K.~N.,  {Bagnulo} S.,
  {Centeno} R.,   {Jes{\'u}s Mart{\'{\i}}nez Gonz{\'a}lez} M.,  eds,  IAU
  Symposium Vol. 305, Polarimetry. pp 61--66

\bibitem[\protect\citeauthoryear{{Niemczura}, {H{\"u}mmerich}, {Castelli},
  {Paunzen}, {Bernhard}, {Hambsch}  \& {He{\l}miniak}}{{Niemczura}
  et~al.}{2017}]{niemczura:2017}
{Niemczura} E.,  {H{\"u}mmerich} S.,  {Castelli} F.,  {Paunzen} E.,  {Bernhard}
  K.,  {Hambsch} F.-J.,   {He{\l}miniak} K.,  2017, \mn@doi [Scientific
  Reports] {10.1038/s41598-017-05987-6}, \href
  {http://adsabs.harvard.edu/abs/2017NatSR...7.5906N} {7, 5906}

\bibitem[\protect\citeauthoryear{{Otero}}{{Otero}}{2003}]{otero:2003}
{Otero} S.~A.,  2003, Information Bulletin on Variable Stars, \href
  {http://adsabs.harvard.edu/abs/2003IBVS.5480....1O} {5480}

\bibitem[\protect\citeauthoryear{{Paunzen} et~al.,}{{Paunzen}
  et~al.}{2018}]{paunzen:2018}
{Paunzen} E.,  et~al., 2018, \aap, in press \href
  {http://adsabs.harvard.edu/abs/2018arXiv180209753P} {} (\mn@eprint {arXiv}
  {1802.09753})

\bibitem[\protect\citeauthoryear{{Perryman} et~al.,}{{Perryman}
  et~al.}{1997}]{perryman:1997}
{Perryman} M.~A.~C.,  et~al., 1997, \aap, \href
  {http://adsabs.harvard.edu/abs/1997A%26A...323L..49P} {323, L49}

\bibitem[\protect\citeauthoryear{{Petit} \& {Wade}}{{Petit} \&
  {Wade}}{2012}]{petit:2012}
{Petit} V.,  {Wade} G.~A.,  2012, \mn@doi [\mnras]
  {10.1111/j.1365-2966.2011.20091.x}, \href
  {http://adsabs.harvard.edu/abs/2012MNRAS.420..773P} {420, 773}

\bibitem[\protect\citeauthoryear{{Pojmanski}}{{Pojmanski}}{2002}]{pojmanski:2002}
{Pojmanski} G.,  2002, Acta Astronomica, \href
  {http://adsabs.harvard.edu/abs/2002AcA....52..397P} {52, 397}

\bibitem[\protect\citeauthoryear{{Pr{\v s}a} \& {Zwitter}}{{Pr{\v s}a} \&
  {Zwitter}}{2005}]{prsa:2005}
{Pr{\v s}a} A.,  {Zwitter} T.,  2005, \mn@doi [\apj] {10.1086/430591}, \href
  {http://adsabs.harvard.edu/abs/2005ApJ...628..426P} {628, 426}

\bibitem[\protect\citeauthoryear{{Pr{\v s}a} et~al.,}{{Pr{\v s}a}
  et~al.}{2016}]{prsa:2016}
{Pr{\v s}a} A.,  et~al., 2016, \mn@doi [\apjs] {10.3847/1538-4365/227/2/29},
  \href {http://adsabs.harvard.edu/abs/2016ApJS..227...29P} {227, 29}

\bibitem[\protect\citeauthoryear{{Renson} \& {Manfroid}}{{Renson} \&
  {Manfroid}}{2009}]{renson:2009}
{Renson} P.,  {Manfroid} J.,  2009, \mn@doi [\aap]
  {10.1051/0004-6361/200810788}, \href
  {http://adsabs.harvard.edu/abs/2009A%26A...498..961R} {498, 961}

\bibitem[\protect\citeauthoryear{{Ros{\'e}n}, {Kochukhov}, {Alecian}, {Neiner},
  {Morin}, {Wade}  \& {the BinaMIcS collaboration}}{{Ros{\'e}n}
  et~al.}{2018}]{rosen:2018}
{Ros{\'e}n} L.,  {Kochukhov} O.,  {Alecian} E.,  {Neiner} C.,  {Morin} J.,
  {Wade} G.~A.,   {the BinaMIcS collaboration} 2018, \aap, \href
  {http://adsabs.harvard.edu/abs/2018arXiv180201031R} {in press}

\bibitem[\protect\citeauthoryear{{Rossiter}}{{Rossiter}}{1924}]{rossiter:1924}
{Rossiter} R.~A.,  1924, \mn@doi [\apj] {10.1086/142825}, \href
  {http://adsabs.harvard.edu/abs/1924ApJ....60...15R} {60, 15}

\bibitem[\protect\citeauthoryear{{Ryabchikova}, {Piskunov}, {Kurucz},
  {Stempels}, {Heiter}, {Pakhomov}  \& {Barklem}}{{Ryabchikova}
  et~al.}{2015}]{ryabchikova:2015}
{Ryabchikova} T.,  {Piskunov} N.,  {Kurucz} R.~L.,  {Stempels} H.~C.,  {Heiter}
  U.,  {Pakhomov} Y.,   {Barklem} P.~S.,  2015, \mn@doi [\physscr]
  {10.1088/0031-8949/90/5/054005}, \href
  {http://adsabs.harvard.edu/abs/2015PhyS...90e4005R} {90, 054005}

\bibitem[\protect\citeauthoryear{{Schneider}, {Podsiadlowski}, {Langer},
  {Castro}  \& {Fossati}}{{Schneider} et~al.}{2016}]{schneider:2016}
{Schneider} F.~R.~N.,  {Podsiadlowski} P.,  {Langer} N.,  {Castro} N.,
  {Fossati} L.,  2016, \mn@doi [\mnras] {10.1093/mnras/stw148}, \href
  {http://adsabs.harvard.edu/abs/2016MNRAS.457.2355S} {457, 2355}

\bibitem[\protect\citeauthoryear{{Sch{\"o}ller}, {Correia}, {Hubrig}  \&
  {Ageorges}}{{Sch{\"o}ller} et~al.}{2010}]{scholler:2010}
{Sch{\"o}ller} M.,  {Correia} S.,  {Hubrig} S.,   {Ageorges} N.,  2010, \mn@doi
  [\aap] {10.1051/0004-6361/201014246}, \href
  {http://adsabs.harvard.edu/abs/2010A%26A...522A..85S} {522, A85}

\bibitem[\protect\citeauthoryear{{Semenko}, {Kudryavtsev}, {Ryabchikova}  \&
  {Romanyuk}}{{Semenko} et~al.}{2008}]{semenko:2008}
{Semenko} E.~A.,  {Kudryavtsev} D.~O.,  {Ryabchikova} T.~A.,   {Romanyuk}
  I.~I.,  2008, \mn@doi [Astrophysical Bulletin] {10.1134/S1990341308020041},
  \href {http://adsabs.harvard.edu/abs/2008AstBu..63..128S} {63, 128}

\bibitem[\protect\citeauthoryear{{Shorlin}, {Wade}, {Donati}, {Landstreet},
  {Petit}, {Sigut}  \& {Strasser}}{{Shorlin} et~al.}{2002}]{shorlin:2002}
{Shorlin} S.~L.~S.,  {Wade} G.~A.,  {Donati} J.-F.,  {Landstreet} J.~D.,
  {Petit} P.,  {Sigut} T.~A.~A.,   {Strasser} S.,  2002, \mn@doi [\aap]
  {10.1051/0004-6361:20021192}, \href
  {http://adsabs.harvard.edu/abs/2002A%26A...392..637S} {392, 637}

\bibitem[\protect\citeauthoryear{{Shultz}, {Wade}, {Alecian}  \& {BinaMIcS
  Collaboration}}{{Shultz} et~al.}{2015}]{shultz:2015}
{Shultz} M.,  {Wade} G.~A.,  {Alecian} E.,   {BinaMIcS Collaboration} 2015,
  \mn@doi [\mnras] {10.1093/mnrasl/slv096}, \href
  {http://adsabs.harvard.edu/abs/2015MNRAS.454L...1S} {454, L1}

\bibitem[\protect\citeauthoryear{{Shultz}, {Rivinius}, {Wade}, {Alecian}  \&
  {Petit}}{{Shultz} et~al.}{2018}]{shultz:2018}
{Shultz} M.,  {Rivinius} T.,  {Wade} G.~A.,  {Alecian} E.,   {Petit} V.,  2018,
  \mn@doi [\mnras] {10.1093/mnras/stx3238}, \href
  {http://adsabs.harvard.edu/abs/2018MNRAS.475..839S} {475, 839}

\bibitem[\protect\citeauthoryear{{Shulyak}, {Tsymbal}, {Ryabchikova},
  {St{\"u}tz}  \& {Weiss}}{{Shulyak} et~al.}{2004}]{shulyak:2004}
{Shulyak} D.,  {Tsymbal} V.,  {Ryabchikova} T.,  {St{\"u}tz} C.,   {Weiss}
  W.~W.,  2004, \mn@doi [\aap] {10.1051/0004-6361:20034169}, \href
  {http://adsabs.harvard.edu/abs/2004A%26A...428..993S} {428, 993}

\bibitem[\protect\citeauthoryear{{Shulyak}, {Ryabchikova}, {Kildiyarova}  \&
  {Kochukhov}}{{Shulyak} et~al.}{2010a}]{shulyak:2010a}
{Shulyak} D.,  {Ryabchikova} T.,  {Kildiyarova} R.,   {Kochukhov} O.,  2010a,
  \mn@doi [\aap] {10.1051/0004-6361/200913750}, \href
  {http://adsabs.harvard.edu/abs/2010A%26A...520A..88S} {520, A88}

\bibitem[\protect\citeauthoryear{{Shulyak}, {Krti{\v c}ka}, {Mikul{\'a}{\v
  s}ek}, {Kochukhov}  \& {L{\"u}ftinger}}{{Shulyak}
  et~al.}{2010b}]{shulyak:2010b}
{Shulyak} D.,  {Krti{\v c}ka} J.,  {Mikul{\'a}{\v s}ek} Z.,  {Kochukhov} O.,
  {L{\"u}ftinger} T.,  2010b, \mn@doi [\aap] {10.1051/0004-6361/201015094},
  \href {http://adsabs.harvard.edu/abs/2010A%26A...524A..66S} {524, A66}

\bibitem[\protect\citeauthoryear{{Sikora}, {Wade}  \& {Power}}{{Sikora}
  et~al.}{2018}]{sikora:2018}
{Sikora} J.,  {Wade} G.~A.,   {Power} J.,  2018, Contributions of the
  Astronomical Observatory Skalnate Pleso, \href
  {http://adsabs.harvard.edu/abs/2018CoSka..48...87S} {48, 87}

\bibitem[\protect\citeauthoryear{{Smith}}{{Smith}}{1996}]{smith:1996}
{Smith} K.~C.,  1996, \mn@doi [\apss] {10.1007/BF02424427}, \href
  {http://adsabs.harvard.edu/abs/1996Ap%26SS.237...77S} {237, 77}

\bibitem[\protect\citeauthoryear{{Strassmeier}, {Granzer}, {Mallonn}, {Weber}
  \& {Weingrill}}{{Strassmeier} et~al.}{2017}]{strassmeier:2017}
{Strassmeier} K.~G.,  {Granzer} T.,  {Mallonn} M.,  {Weber} M.,   {Weingrill}
  J.,  2017, \mn@doi [\aap] {10.1051/0004-6361/201629150}, \href
  {http://adsabs.harvard.edu/abs/2017A%26A...597A..55S} {597, A55}

\bibitem[\protect\citeauthoryear{{Valyavin}, {Kochukhov}  \&
  {Piskunov}}{{Valyavin} et~al.}{2004}]{valyavin:2004}
{Valyavin} G.,  {Kochukhov} O.,   {Piskunov} N.,  2004, \mn@doi [\aap]
  {10.1051/0004-6361:20034345}, \href
  {http://adsabs.harvard.edu/abs/2004A%26A...420..993V} {420, 993}

\bibitem[\protect\citeauthoryear{{Vick}, {Michaud}, {Richer}  \&
  {Richard}}{{Vick} et~al.}{2010}]{vick:2010}
{Vick} M.,  {Michaud} G.,  {Richer} J.,   {Richard} O.,  2010, \mn@doi [\aap]
  {10.1051/0004-6361/201014307}, \href
  {http://adsabs.harvard.edu/abs/2010A%26A...521A..62V} {521, A62}

\bibitem[\protect\citeauthoryear{{Woolf} \& {Lambert}}{{Woolf} \&
  {Lambert}}{1999}]{woolf:1999a}
{Woolf} V.~M.,  {Lambert} D.~L.,  1999, \mn@doi [\apjl] {10.1086/312146}, \href
  {http://adsabs.harvard.edu/abs/1999ApJ...520L..55W} {520, L55}

\makeatother
\end{thebibliography}

\appendix

\section{Details of {\sc PHOEBE} analysis}
\label{phoebe}

To obtain a {\sc PHOEBE} model with robustly determined errors we employed a Bayesian Markov Chain Monte Carlo (MCMC) numerical scheme. Furthermore, to determine uncertainties accounting for the effect of known correlations between binary model parameters, we wrapped {\sc PHOEBE} into the {\sc EMCEE} code by \citet{foreman-mackey:2013}, which makes use of an ensemble, affine-invariant approach to sampling the parameter posterior distributions.

The MCMC optimisation is based upon Bayes' Theorem: 
\begin{equation}
p\left(\Theta|d\right)\propto\mathcal{L}\left(d|\Theta\right)\,p\left(\Theta\right).
\label{bayes}
\end{equation}
Here, $p\left(\Theta|d\right)$ is the posterior distribution of the model parameters $\Theta$, which are varied in the optimisation given the observations $d$. The prior probability of the parameter vector $p\left(\Theta\right)$ is used to draw the parameter configurations, and is set as either uniform or Gaussian. The likelihood function $\mathcal{L}\left(\Theta|d\right)$ evaluates the likelihood that a given parameter configuration accurately reproduces the observations. For computational ease, we work in log-space where we write the likelihood as:
\begin{equation}
\label{likelihood}
\mathrm{ln}\mathcal{L}\propto -\frac{1}{2}\sum_{i}\left( \frac{d_i - M\left(\Theta\right)_i}{\sigma_i} \right)^2,
\end{equation} 
where $\sigma_i$ represent the uncertainties on each data point $d_i$  and $M\left( \Theta \right)_i$ is the model produced using the parameter configuration $\Theta$.

For the MCMC analysis
we used an ensemble configuration of 138 individual parameter chains and allowed the algorithm to run for 5000 iterations. After 5000 iterations, we checked the chains for convergence via auto-correlation. Once the samplings have been confirmed to have converged, we burned the first 1500 iterations and produced the posteriors for the remaining 3500 iterations. We calculated 95\% confidence intervals (CIs) by evaluating the Highest Posterior Density (HPD) with $\alpha=0.05$ for all marginalised posteriors. We varied 14 parameters in total, 4 of which are passband luminosities related to scaling the model light curves to the observations, while the remaining 10 are physical parameters, all of which are listed in Table~\ref{parameters_table}. 

\begin{figure}
\centering
\figps{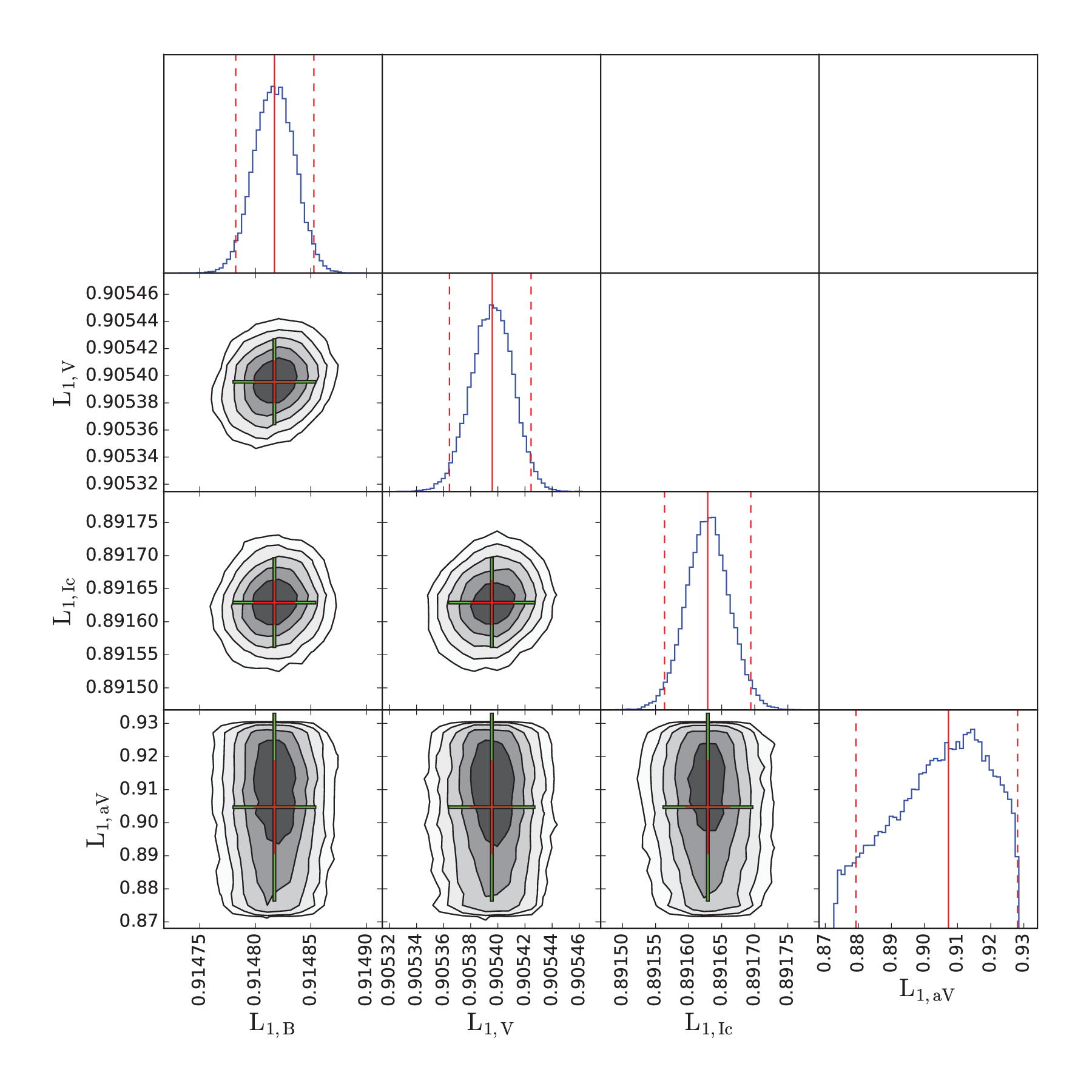}
\caption{Marginalised Posterior Distributions for primary light fraction for each observed filter. Median denoted by solid vertical red line, upper and lower bounds for 95\% CI denoted by dashed vertical red lines.}
\label{fig:luminosity_posteriors}
\end{figure}

\begin{figure}
\centering
\figps{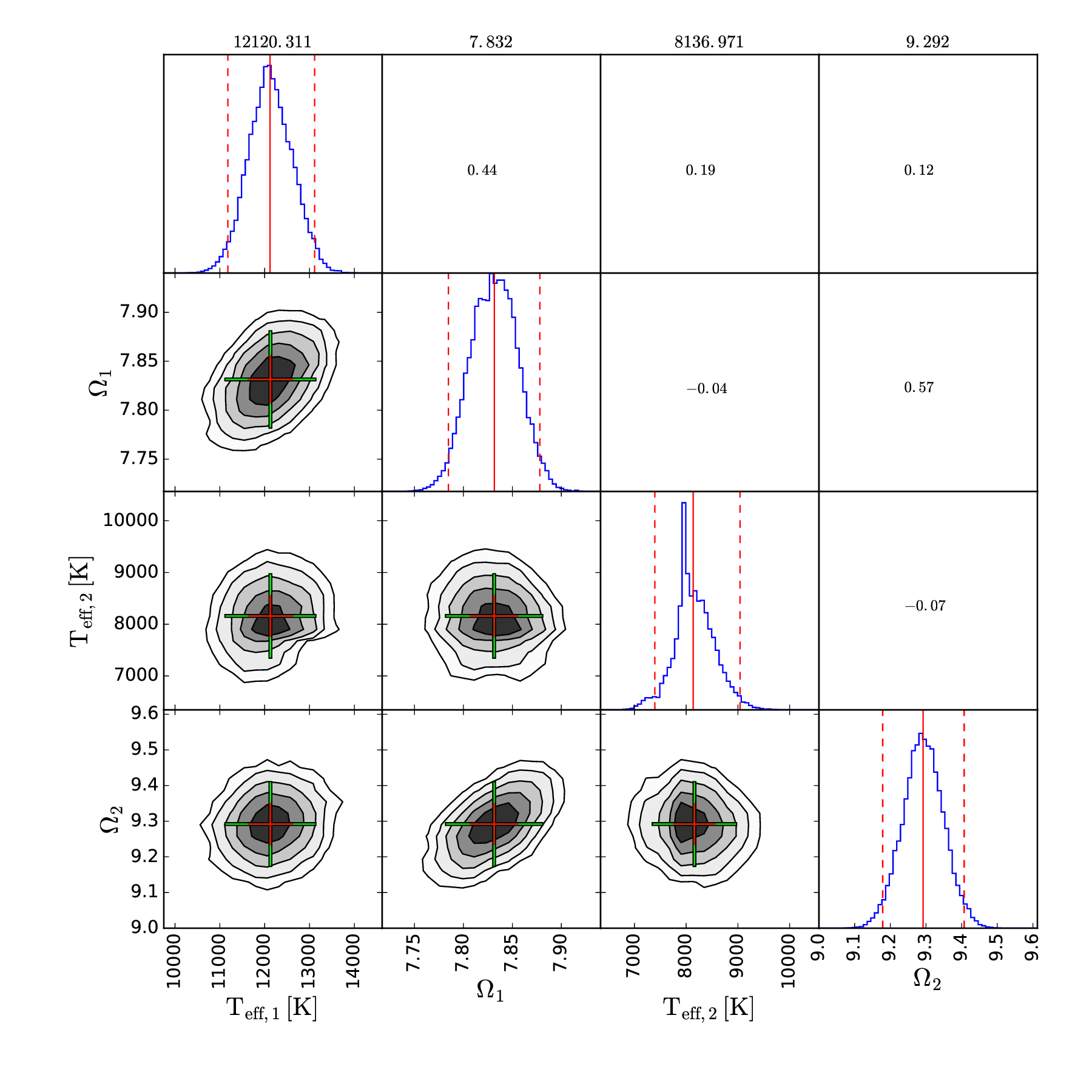}
\caption{Marginalised Posterior Distributions for primary and secondary parameters. Median denoted by solid vertical red line, upper and lower bounds for 95\% CI denoted by dashed vertical red lines.}
\label{fig:star_posteriors}
\end{figure}

\begin{figure}
\centering
\figps{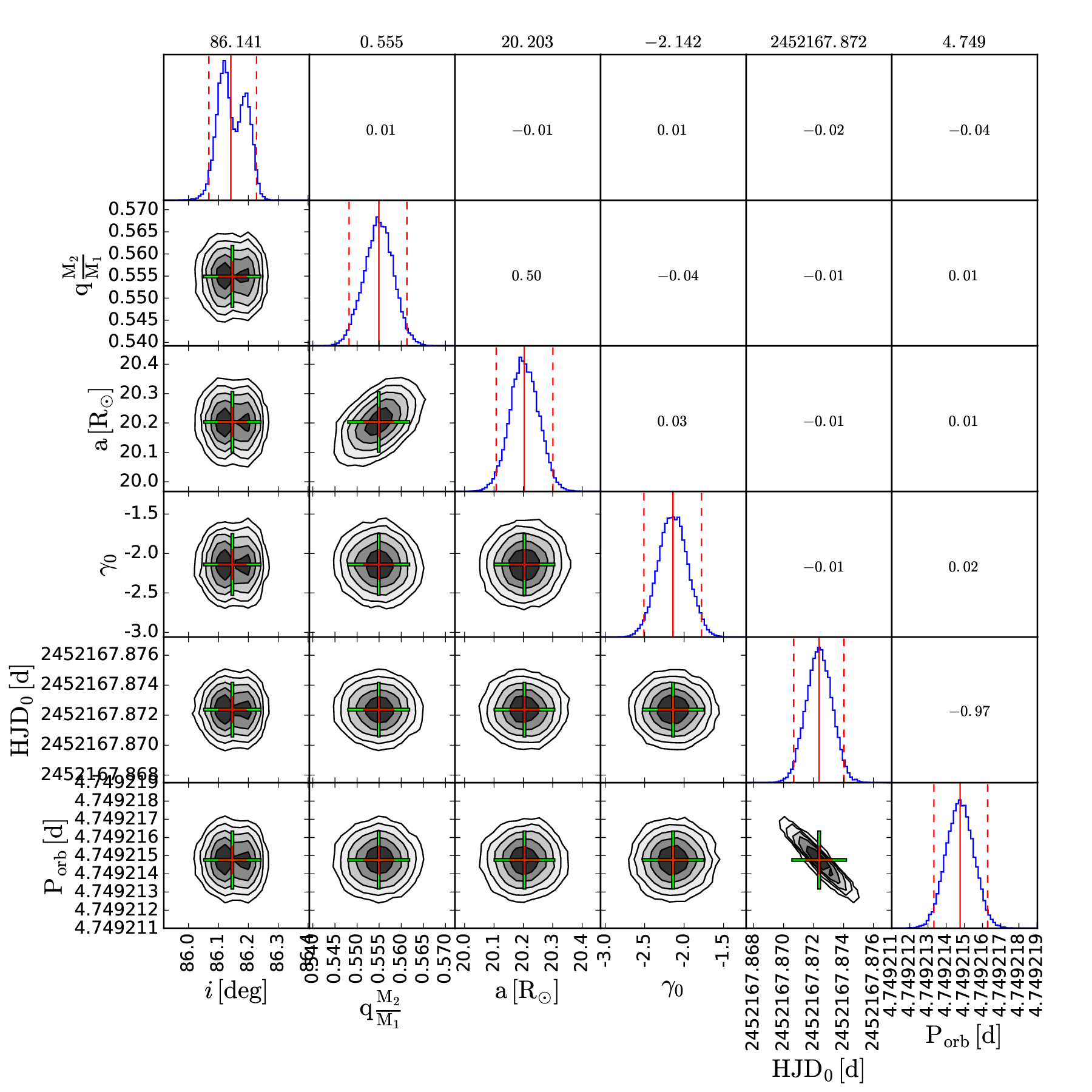}
\caption{Marginalised Posterior Distributions for system parameters. Median denoted by solid vertical red line, upper and lower bounds for 95\% CI denoted by dashed vertical red lines.}
\label{fig:system_posteriors}
\end{figure}

The marginalised posteriors and HPD error estimates are shown in Figs.~\ref{fig:luminosity_posteriors} to \ref{fig:system_posteriors}. Figure~\ref{fig:luminosity_posteriors} shows the marginalised posteriors for the passband luminosities derived by {\sc PHOEBE}. It is clear from these posteriors that the ASAS $V$-band curve has little impact on the likelihood as the posteriors recover the priors. This can be understood in terms of the larger formal error bars associated with these data compared to the other light curves. Figures~\ref{fig:star_posteriors} and \ref{fig:system_posteriors} show the posteriors for individual stellar parameters and system parameters, respectively. While most posteriors are well behaved, the posteriors for the inclination, $i$, show complex structures, best described as an unresolved bimodal distribution. However, as the HPD estimation includes both nodes in the CIs, the errors are propagated properly. This is reflected in the propagated errors on the derived stellar masses in Table~\ref{derived_table}, however, it does not push the  estimate of either mass more than a few percent above or below their median values. Furthermore, despite a Gaussian prior on $T_{\rm eff}^{(2)}$, the posteriors climbed to systematically higher values, and has a CI range of 1600 K. 

The final light curve and radial velocity variation models presented in the paper were constructed using the median values of the parameters estimated from marginalised posteriors.

\bsp
\label{lastpage}

\end{document}